\documentclass[journal]{IEEEtran}
\IEEEoverridecommandlockouts
\usepackage[noadjust]{cite}
\usepackage[hidelinks,pageanchor=false,hypertexnames=false]{hyperref}
\usepackage{booktabs}
\usepackage{amsmath,amssymb,amsfonts,bbm}
\usepackage{dsfont}
\usepackage{algpseudocode}
\usepackage{algorithm}
\algrenewcommand\algorithmicfunction{}
\usepackage{graphicx}
\usepackage{epstopdf}
\usepackage{textcomp}
\usepackage[dvipsnames]{xcolor}
\usepackage{comment}
\usepackage{mathtools}
\usepackage{ifthen}
\usepackage[capitalize]{cleveref}
\crefrangeformat{line}{Lines~#3#1#4--#5#2#6}
\usepackage[caption=false]{subfig}
\renewcommand{\Cref}[1]{\cref{#1}}

\newtheorem{lemma}{Lemma}
\newtheorem{proposition}{Proposition}

\DeclareMathOperator*{\minimize}{minimize}

\usepackage{xparse}

\begin{document}

\title{Black-Box Edge AI Model Selection with\\Conformal Latency and Accuracy Guarantees}

\author{Anders~E.~Kalør,~\IEEEmembership{Member,~IEEE}, and Tomoaki Ohtsuki,~\IEEEmembership{Senior Member,~IEEE}%
  \thanks{The work of A.~E.~Kal{\o}r and T.~Ohtsuki was supported in part by the Japan Science and Technology Agency (JST) ASPIRE program (grant no. JPMJAP2326). The work A.~E.~Kal{\o}r was also supported in part by the Mizuho Foundation for the Promotion of Sciences.}%
  \thanks{A.~E.~Kal{\o}r and T.~Ohtsuki are with the Department of Information and Computer Science, Keio University, Yokohama 223-8522, Japan. (e-mails: \{aek, ohtsuki\}@keio.jp).}%
}

\maketitle
\begin{abstract}
  Edge artificial intelligence (AI) will be a central part of 6G, with powerful edge servers supporting devices in performing machine learning (ML) inference. However, it is challenging to deliver the latency and accuracy guarantees required by 6G applications, such as automated driving and robotics. This stems from the black-box nature of ML models, the complexities of the tasks, and the interplay between transmitted data quality, chosen inference model, and the random wireless channel. This paper proposes a novel black-box model selection framework for reliable real-time wireless edge AI designed to meet predefined requirements on both deadline violation probability and expected loss. Leveraging conformal risk control and non-parametric statistics, our framework intelligently selects the optimal model combination from a collection of black-box feature-extraction and inference models of varying complexities and computation times. We present both a fixed (relying on channel statistics) and a dynamic (channel-adaptive) model selection scheme. Numerical results validate the framework on a deadline-constrained image classification task while satisfying a maximum misclassification probability requirement. These results indicate that the proposed framework has the potential to provide reliable real-time edge AI services in 6G.
\end{abstract}


\section{Introduction}
Driven by the success of artificial intelligence (AI), edge AI is expected to be a central component of 6G, where servers located at the edge of the network will support devices in performing inference and making decisions using machine learning (ML)~\cite{letaief2021edgeai,massiveandcritical24}. For instance, powerful edge servers may assist vehicles in performing image object detection in automated driving~\cite{zhang20vehicularedge}, or execute complex reinforcement learning models to control industrial robots~\cite{knowledgebasedurllcinf}. Such edge AI applications often operate under strict performance and time constraints, requiring inference results to be both accurate and delivered before a specific deadline with high probability. For instance, an augmented reality application in a factory setting may demand an end-to-end latency less than 20 milliseconds and reliability in the order of $1-10^{-5}$~\cite{3gpp_ts22104}.

Meeting these requirements involves inherent trade-offs between the quality of transmitted data representations (affecting accuracy and uplink time), the computational complexity of edge ML models (affecting accuracy and processing time), and the size of the resulting predictions (affecting downlink transmission). For example, in an image classification scenario the device may first compress its image using a lossy compression algorithm with an adjustable quality parameter that controls the trade-off between the distortion and the size of the compressed image. The quality of the transmitted image in turn influences the inference quality at the edge server. Similarly, the edge server may have access to an ensemble of classifier models, each of various size and accuracy, so that a large model is more likely to produce accurate predictions but has a longer computation time~\cite{pmlr-v97-tan19a}. Finally, the prediction accuracy may influence the number of produced class labels (e.g., top-1, top-5, top-10) required to meet the desired accuracy.
Directly optimizing this trade-off is challenging due to the interaction between data quality and model accuracy, which depends on the specific prediction task and is hard to quantify, combined with the stochastic nature of wireless channels.

However, the problem of providing provable performance guarantees for ML models has recently seen significant advancements through the application of non-parametric statistics~\cite{conformal,conformalrisk,angelopoulosppi}. Conformal risk control, in particular, offers a powerful and promising tool for achieving distribution-free performance guarantees for black-box ML models~\cite{conformalrisk}.
Building upon non-parametric statistics and conformal risk control, in this paper we propose a generalized framework for black-box model selection that provides strict statistical guarantees on the resulting end-to-end loss and latency. The framework directly accounts for challenges such as variable compression rates and interactions between compression settings and model accuracy. Given a loss function and a deadline, our framework intelligently chooses the best transmission quality and edge inference model, by selecting from ensembles of black-box models, to guarantee that the final predictions satisfy predefined requirements for loss and latency (see \cref{fig:scenario}). The key principle behind our method is to statistically bound the loss and delay of each model using a \emph{calibration} dataset, and then selecting the best model combination among the subset of combinations that meet the requirements.

\subsection{Related Work}
Several works have studied and optimized the trade-off between latency and accuracy in wireless edge AI. A popular technique is split inference~\cite{shao2020edgeai,huang2020edge,Chen-TWC-2019,li2028jalad,jankowski2020edgeinf,jankowski2021imgretrieval,knowledgebasedurllcinf,urllc-inference,urllc-sensing}, in which a neural network is vertically split into two parts. The initial layers are executed by the device followed by transmitting the intermediate layer representations to the edge server, which executes the final layers of the model. Similar techniques include early exiting~\cite{liu23earlyexiting,she23edgescheduling,jankowski2024earlyexit}, where the neural network is terminated at intermediate layers if the inference confidence is sufficiently high, and over-the-air computing~\cite{Zhiyan-AirPooling}, where the linearity of the wireless medium is exploited for fast multi-view inference. Another line of work focuses on feature transmission for low-latency edge AI, e.g., by through progressive feature transmission~\cite{lan23progressive}, or by considering finite-blocklength effects~\cite{urllc-inference}. Common to these methods is that they rely on \emph{white-box} ML models, and their analysis either rely on oversimplified data models, which rarely reflect practical settings, or focuses on aggregate performance metrics such as average accuracy and latency, which is insufficient for critical applications. On the other hand, many state-of-the-art ML models, including large language models (LLMs) and diffusion models, are either provided \emph{as a service}~\cite{zhang25beyondcloud} and inherently black-box, or too complex for white-box analysis.
Furthermore, resource-constrained devices may not be able to compute complex features as required by split inference, relying instead on simple processing tools, such as image compression with various quality settings. Thus, while the aforementioned methods aim at reducing inference latency in edge AI scenarios, they are insufficient in providing the end-to-end performance guarantees required in 6G. Our framework addresses these shortcomings by providing rigorous end-to-end guarantees under a black-box assumption, capturing both ML-based feature extraction as in traditional split inference and more classical source coding techniques, such as standard image compression algorithms.

Conformal risk control, and the closely related conformal prediction framework~\cite{conformal}, have previously been successfully applied to ML problems in the context of wireless communication, including scheduling~\cite{cohen23conformalurllc}, channel prediction and modulation detection~\cite{cohen23conformal}, and federated learning~\cite{zhu24federated}.
The key idea behind conformal risk control is to output a \emph{set of predictions} rather than a single point estimate. Through careful construction and calibration of the prediction set, conformal risk control ensures that the expected loss of unseen examples is upper bounded by a specified constant, thereby providing reliable predictions and quantifiable uncertainty estimates. By explicitly quantifying uncertainty, conformal risk control enables edge AI applications to make more nuanced decisions based on the confidence level associated with each prediction, thereby enhancing reliability and robustness.
However, unlike traditional point predictions which output a single prediction, the size of the prediction set produced with conformal risk control is random and depends on the uncertainty of with the prediction task and the desired confidence level. This necessitates a joint design of the communication subsystem and the prediction model that accounts for both the model computation time, the reliability of the predictions, and the communication of the resulting prediction sets to ensure that the results are both reliable and delivered before the deadline.

\subsection{Main Contributions}
In this paper, we apply the conformal risk control framework and non-parametric statistics to address the problem of black-box edge AI model selection under strict end-to-end reliability and deadline requirements. The main contributions of this work can be summarized as follows:

\begin{itemize}
\item We propose a novel framework for providing statistically sound, end-to-end performance guarantees for black-box edge AI systems. This is achieved through a novel integration of conformal risk control to meet a pre-defined expected loss requirement, with non-parametric statistics to bound the deadline violation probability, explicitly accounting for variable message lengths and random wireless channel conditions.
\item We develop a fixed and a dynamic model selection scheme. The fixed scheme optimizes the combination of observation encoder/decoder and edge inference model a priori based on channel statistics. The dynamic scheme adapts the choice of the edge inference model based on the instantaneous uplink channel conditions, after an initial encoder/decoder selection. Both schemes aim to minimize the prediction set size while satisfying specified loss and deadline guarantees.
\item We demonstrate and validate the effectiveness of the proposed framework on a realistic deadline-constrained image classification task using standard datasets and models. The results show that the proposed schemes can successfully meet stringent requirements on maximum misclassification probability and deadline violation, while adapting model choices to channel quality.
\end{itemize}

The remainder of the paper is organized as follows. \Cref{sec:sysmodel} introduces the system model and formalizes the problem. The conformal risk control framework, which we apply to provide reliable black-box predictions, is described in \cref{sec:conformalrisk}. \Cref{sec:fixed_selection} and \cref{sec:dynamic_selection} present the proposed fixed and dynamic model selection schemes, respectively. The framework is validated through numerical results in \cref{sec:numres}, and finally the paper is concluded in \cref{sec:conclusion}.

\begin{figure*}
  \centering
  \includegraphics{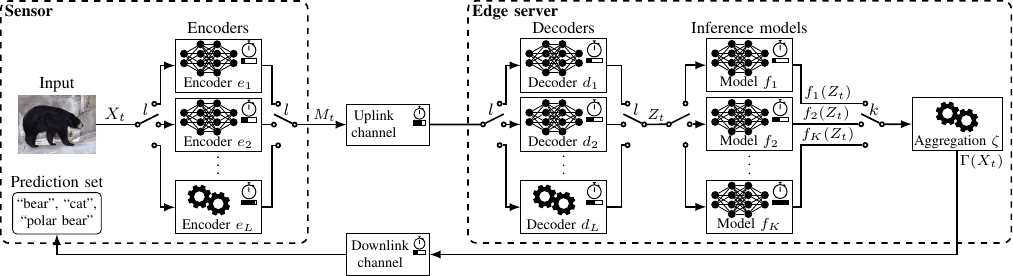}
  \caption{The considered edge inference scenario. The sensor encodes its input $X_t$ using one of its $L$ encoders and transmits it to the edge server. The server decodes the received observation to an intermediate representation $Z_t$ using its corresponding decoder, and then performs inference using one of its $K$ inference models. The inference model outputs are then aggregated into a prediction set $\Gamma(X_t)$, which is transmitted back to the sensor.}\label{fig:scenario}
\end{figure*}

\section{System Model and Problem Statement}\label{sec:sysmodel}
We consider the system depicted in \cref{fig:scenario}, in which a deadline-constrained sensor is connected to a single edge server over a wireless link. Time is divided into frames of a fixed duration $T$ seconds, indexed by $t=1,2,\ldots$. In each frame, the sensor observes an input $X_t\in\mathcal{X}$, on which it wishes to perform real-time inference, such as classification or image object detection, assisted by the edge server. Associated with the input $X_t$ is a set of \emph{unobservable} ground-truth labels $Y_t\subseteq\mathcal{Y}$, where $\mathcal{Y}$ is a discrete, finite (but possibly large) set comprising all possible labels. For instance, $X_t$ could be an image and $Y_t$ could be all objects in the image, possibly along with bounding boxes defined on the pixels. We assume that $(X_t,Y_t)$ are independent across frames and drawn from an unknown joint distribution $P_{XY}$. To simplify the notation, we will omit the temporal dependence on $t$ unless it is essential. Owing to a strict end-to-end deadline constraint, the entire inference task, comprising the transmission of the input, edge inference, and transmission of the result in the downlink, must be completed before the end of the frame. Next, we describe the various phases in detail and present the overall objective.

\subsection{Observation Transmission}
The observed input $X_t$ is encoded into a binary message that can be transmitted to the edge server over the channel. To this end, we assume that the sensor and the edge server have access to a collection of $L$ encoder/decoder pairs $(e_1,d_1),\ldots,(e_L,d_L)$, each defined as
\begin{align*}
  &e_l: \mathcal{X}\to \{0,1\}^{*},\\
  &d_l: \{0,1\}^{*}\to\mathcal{Z},
\end{align*}
for $l=1,2,\ldots,L$, where $\{0,1\}^{*}$ denotes the set of all finite-length binary strings. In words, each encoder $e_l$ produces a message $M_t=e_{l}(X_t)$ of (variable) length $D_{\mathrm{ul},l}(X_t)$ bits, while the corresponding decoder $d_l$ decodes $M_t$ into an intermediate representation $Z_t=d_l(M_t)\in\mathcal{Z}$. The intermediate representation $Z_t$ will be used for subsequent inference at the edge server, and is assumed to belong to some common space $\mathcal{Z}$ that is common to all encoder/decoder pairs and serves as the interface between the different encoder/decoder pairs and the edge inference models\footnote{Although we assume that each encoder/decoder pair can interface to any edge inference model, it is straightforward to apply our framework to the more general case where each encoder/decoder pair only supports a subset of the edge models.}. We assume that the $l$-th encoder/decoder pair has a fixed and deterministic total computation time $\tau_{\mathrm{ul},l}$, comprising both encoding and decoding but excluding transmission delay. Thus, each pair offers a different trade-off between the size of the encoded message $D_{\mathrm{ul},l}$ (and hence the communication delay), the encoding/decoding computation time $\tau_{\mathrm{ul},l}$, and the fidelity of the intermediate representation $Z_t$ decoded by the edge server. For instance, one encoder/decoder pair $(e_l,d_l)$ might have a high compression ratio, resulting in a small message size, but be slow and lead to an imprecise reconstruction at the edge server. Conversely, another pair might use a lower compression ratio, leading to a longer message but a more accurate representation and a low computation time. 

In each frame $t=1,2,\ldots$, a single encoder/decoder pair $(e_{l_t},d_{l_t})$ is used for transmitting the input $X_t$. The data transmission happens over a quasi-static fading wireless channel with additive white Gaussian noise (AWGN), in which the channel remains constant throughout the transmission and changes independently between transmissions. The communication rate is given by
\begin{equation*}
  R_{\mathrm{ul},t} = B\log_2\left(1+|h_{\mathrm{ul},t}|^2\mathsf{SNR}_{\mathrm{ul}}\right)\ \ \text{(bits/s)}, \label{eq:capacity_ul}
\end{equation*}
where $B$ is the bandwidth in Hz, $h_{\mathrm{ul},t}\sim\mathcal{CN}(0,1)$ is the instantaneous uplink channel coefficient in frame $t$, and $\mathsf{SNR}_{\mathrm{ul}}$ is the average signal-to-noise ratio (SNR), which is known to both the device and the edge server. The total duration of the observation transmission can then be computed as
\begin{equation}
  T_{\mathrm{ul},t} = \tau_{\mathrm{ul},l_t} + \frac{D_{\mathrm{ul},l_t}(X_t)}{R_{\mathrm{ul},t}}.\label{eq:T_ul}
\end{equation}
Motivated by the fact that channel state information (CSI) may not be available at the application layer and the potentially long encoding time, we assume that only the statistics of $R_{\mathrm{ul},t}$, and not the instantaneous realization, can be used to select the encoder/decoder pair.

\subsection{Edge Inference and Result Transmission}
After successful transmission of the message $M_t$, the edge server performs inference on the decoded intermediate representation $Z_t=d_l(M_t)\in\mathcal{Z}$. We assume that the edge server has access to $K$ pre-trained, black-box inference models $\{f_k\}_{k=1}^K$.
Each inference model $f_k$, $k=1,\ldots,K$, takes the representation $Z_t$ as input (regardless of the encoder/decoder pair used for transmission) and outputs a confidence score $[f_k(Z_t)]_y$ of each label $y\in\mathcal{Y}$, e.g., using the softmax activation function. We do not impose any assumptions on how these models are trained, and they could be trained on datasets that follow a different distribution than $P_{XY}$.
The $k$-th model has a fixed computation time $\tau_{f_k}$, and, without loss of generality, we assume $\tau_{f_1}\le \tau_{f_2}\le\ldots\le \tau_{f_K}$. Typically, a model with a longer computation time is expected to produce better predictions, but this may not always be the case. The inference models could, for instance, be implemented as a single neural network with $K-1$ ``early exits''~\cite{earlyexiting}, where each exit point produces an output before the final layers of the network, or by having multiple \emph{scales} of the same model, each with a different number of layers, neurons, etc.~\cite{wang2024yolov10}. However, we emphasize that the considered edge AI model is general and agnostic to the specific architecture of the underlying ML models, treating them effectively as black boxes.

In each frame, the edge server selects and executes one of its inference models $f_{k_t}$, $k_t\in\{1,\ldots,K\}$, to obtain a set of confidence scores $\{[f_{k_t}(Z_t)]_y\,:\,y\in\mathcal{Y}\}$.
Using these, the edge server constructs a \emph{prediction set} $\Gamma(X_t)\subseteq\mathcal{Y}$ by applying an aggregation function $\zeta$ to the set of confidence scores produced by the selected model:
\begin{equation}
  \Gamma(X_t)=\zeta\left(\left\{[f_{k_t}(Z_t)]_y \,:\, y\in\mathcal{Y}\right\}\right).\label{eq:zeta_def}
\end{equation}
Note that the prediction set $\Gamma(X_t)$ contains a set of labels rather than a single point estimate. For instance, $\zeta$ could select the $\kappa$ labels with the largest confidence scores, or all labels with a confidence score greater than some threshold.
In general, the choice of aggregation function $\zeta(\cdot)$ controls the trade-off between \emph{coverage} (i.e., the fraction of contained ground-truth labels) and \emph{informativeness} (i.e., the size of the prediction set). The specific implementation of $\zeta$ will be detailed later. Note also that $\Gamma(X_t)$ is a random quantity that depends on the random input $X_t$ through $Z_t$, which in turn is affected by the random uplink channel.

As in the uplink, the prediction set is transmitted back to the sensor over a channel with rate
\begin{equation}
  R_{\mathrm{dl},t} = B\log_2\left(1+|h_{\mathrm{dl},t}|^2\mathsf{SNR}_{\mathrm{dl}}\right)\ \ \text{(bits/s)}, \label{eq:capacity_dl}
\end{equation}
where the average SNR $\mathsf{SNR}_{\mathrm{dl}}$ is assumed to be known, while the instantaneous fading coefficient $h_{\mathrm{dl},t}\sim\mathcal{CN}(0,1)$ is revealed to the edge server only prior to communication.
To this end, we assume that each predicted label $y\in\Gamma(X_t)$ occupies $D_{\mathrm{lbl}}$ bits, so that the transmitted prediction set can be represented by
\begin{equation*}
  D_{\mathrm{dl},l_t,k_t}(X_t)=|\Gamma(X_t)|D_{\mathrm{lbl}}\ \ \text{(bits)}.
\end{equation*}
Since each label may include metadata such as bounding box coordinates, depth estimates, textual descriptions, etc., $D_{\mathrm{lbl}}$ can potentially span from a few bits to several hundred bytes depending on the application.
The edge inference and downlink transmission time is thus given as
\begin{equation}
  T_{\mathrm{dl},t} = \tau_{f_{k_t}} + \frac{D_{\mathrm{dl},l_t,k_t}(X_t)}{R_{\mathrm{dl},t}}.\label{eq:T_dl}
\end{equation}
Throughout the paper, we will assume that the transmission is terminated when the frame ends, so that a transmission error occurs whenever $T_{\mathrm{dl},t}>T$.

\subsection{Metrics and Problem Statement}\label{sec:metrics_problem_statement}
Given the critical nature of the prediction task, we are interested in generating an accurate prediction set $\Gamma(X_t)$ in each frame that can be delivered to the sensor within the frame duration $T$. To this end, we assume that the quality of a prediction set is characterized by a loss function $\ell(\Gamma(X_t), Y_t)$ that quantifies how well the predictions $\Gamma(X_t)$ correspond to the ground-truth labels $Y_t\subseteq\mathcal{Y}$, so that a good prediction set yields a low loss. For technical reasons, we assume that the loss can never increase by enlarging $\Gamma(X_t)$, and that it is upper bounded by some constant $\gamma$. Note that these conditions are satisfied for many common loss functions, such as the 0--1 loss and the false negative rate. We are interested in producing prediction sets that ensure the expected loss for a test sample $(X,Y)\sim P_{XY}$ is at most $\alpha$, i.e.,
\begin{equation}
  \mathbb{E}_{(X,Y)\sim P_{XY}}[\ell(\Gamma(X),Y)]\le \alpha.\label{eq:conformalrisk}
\end{equation}
Note that by using the indicator function as the loss function, the expectation in \cref{eq:conformalrisk} becomes equivalent to a probability, and thus the expression can be used to bound, e.g., the probability of a false negative prediction.

The requirement in \cref{eq:conformalrisk} can be satisfied by simply including all labels in $\mathcal{Y}$ (or a random fraction $\alpha$ of the labels). However, this solution would obviously be completely uninformative. Instead, we aim to return the \emph{smallest} prediction set satisfying~\eqref{eq:conformalrisk}. Specifically, we seek to design a procedure to select the observation encoder/decoder, the edge inference model, and the aggregation function $\zeta$, such that the size of any received prediction set $\Gamma(X_t)$ is minimized while satisfying \cref{eq:conformalrisk} and having a missed deadline probability of at most $\beta$. This can be formally stated as:
\begin{subequations}\label{eq:problemdef}
  \begin{align}
    \minimize \quad & \mathbb{E}\left[|\Gamma(X_t)|\,\mid\,T_{\text{tot},t}\le T\right],\label{eq:problemdefobj}\\
    \textrm{s.t.} \quad & \mathbb{E}\left[\ell\left(\Gamma(X_t),Y_t\right)\,\mid\,T_{\text{tot},t}\le T\right] \le \alpha,\label{eq:problemdefstconformal}\\
    & \Pr\left(T_{\text{tot},t}> T\right) \le \beta,\label{eq:problemdefstoutage}
  \end{align}
\end{subequations}
where $T_{\mathrm{tot},t}=T_{\mathrm{ul},t}+T_{\mathrm{dl},t}$, and the expectations and the probability are over both $(X_t,Y_t)\sim P_{XY}$ and $h_{\mathrm{ul},t},h_{\mathrm{dl},t}\sim\mathcal{CN}(0,1)$.

Solving the problem in~\eqref{eq:problemdef} optimally is generally challenging since $P_{XY}$ is unknown. Instead, we assume access to labeled and unlabeled \emph{calibration} datasets. The labeled dataset is denoted by $\mathcal{D}$ and contains $N_{\mathcal{D}}$ samples drawn independently and identically distributed (i.i.d.) from $P_{XY}$, i.e.,
\begin{equation*}
  \mathcal{D}=\{(X_n^{(\mathcal{D})},Y_n^{(\mathcal{D})})\}_{n=1}^{N_{\mathcal{D}}},\quad (X_n^{(\mathcal{D})},Y_n^{(\mathcal{D})}) \overset{\text{i.i.d.}}{\sim} P_{XY}.
\end{equation*}
Similarly, the unlabeled dataset, denoted by $\mathcal{U}$, contains $N_{\mathcal{U}}$ i.i.d.\ samples from the marginal input distribution, $P_X$, of $P_{XY}$:
\begin{equation*}
  \mathcal{U}=\{X_n^{(\mathcal{U})}\}_{n=1}^{N_{\mathcal{U}}},\quad X_n^{(\mathcal{U})} \overset{\text{i.i.d.}}{\sim} P_{X}.
\end{equation*}
Utilizing these datasets, our aim is to devise model selection procedures that are guaranteed to satisfy the loss and deadline constraints on \emph{unseen samples} drawn from $P_{XY}$, while having small, but not necessarily minimal, prediction sets.

\section{Reliable Edge Predictions through Conformal Risk Control}\label{sec:conformalrisk}
In this section, we present the conformal risk control framework~\cite{conformalrisk} and show how it can be used to design the aggregation function $\zeta$ in \cref{eq:zeta_def} to ensure that constraint~\eqref{eq:problemdefstconformal} is satisfied for any combination of encoder/decoder model and inference model.

Conformal risk control, belonging to the conformal prediction framework~\cite{conformal}, is a tool for providing model-agnostic and distribution-free statistical guarantees for ML model predictions. It leverages the calibration dataset to provide a model calibration framework. Specifically, conformal risk control enables us to select which predictions to include in the transmitted prediction set based on the confidence scores produced by the executed model $f_{k_t}$, so that the expected loss requirement in \cref{eq:problemdefstconformal} is met.
To keep the presentation clear, we defer the discussion of the impact of communication constraints on model selection and performance to \cref{sec:fixed_selection,sec:dynamic_selection}.

To understand conformal risk control, it is instructive to first consider a single model $f$ that takes as input directly the input $X\in\mathcal{X}$ and outputs a confidence score $[f(X)]_y$ for each label $y\in\mathcal{Y}$. In our scenario, this corresponds to using a lossless encoder/decoder pair that outputs an intermediate representation $Z\in\mathcal{Z}$ that is equal to $X$.
The operating principle behind conformal risk control is to select the aggregation function $\zeta$ that outputs all predictions whose confidence score exceeds a fixed threshold $1-\lambda$, i.e., constructing the prediction set as
\begin{align}
  \Gamma(X) &=\left\{y\in\mathcal{Y}\, :\, [f(X)]_y \ge 1-\lambda\right\}.\label{eq:conformalpredictionset}
\end{align}
Note that a large $\lambda$ includes more items into the prediction set, leading to a smaller loss but less informative prediction set.

Our main task is to use the labeled calibration dataset $\mathcal{D}$ to select the smallest $\lambda$ such that the expected loss, taken over the \emph{true distribution} $P_{XY}$ is guaranteed to be no larger than some fixed constant $\varepsilon$.
As formally stated in the following lemma, this can be achieved by selecting the threshold as the quantile of the empirical confidence score distribution while correcting for the finite size of the calibration dataset.
\begin{lemma}[Conformal risk control~\cite{conformal,conformalrisk}]\label{lemma:conformalrisk}
  Let $\mathcal{D}=\{(X_n^{(\mathcal{D})},Y_n^{(\mathcal{D})})\}_{n=1}^{N_{\mathcal{D}}}$ be a set of $N_{\mathcal{D}}$ samples drawn i.i.d. from $P_{XY}$, and let $\Gamma_{\lambda}(x)$ denote the prediction set constructed from input $x$ using \cref{eq:conformalpredictionset} for a fixed model $f$ with the threshold $\lambda$. Suppose the loss function $\ell$ satisfies
  \begin{align}
    \ell(\Gamma_{\lambda_2}(x), y)\le \ell(\Gamma_{\lambda_1}(x), y) \le \gamma \label{eq:conformal_loss_assumption}
  \end{align}
  for all $(x,y)$ and $\lambda_1\le \lambda_2$, and for some finite $\gamma$.
  The threshold
  \begin{align}
    \lambda^{*}=\inf\left\{\lambda \,:\, \frac{1}{N_{\mathcal{D}}}\sum_{n=1}^{N_{\mathcal{D}}} \ell(\Gamma_{\lambda}(X_n^{(\mathcal{D})}), Y_n^{(\mathcal{D})}) \le \varepsilon-\frac{\gamma-\varepsilon}{N_{\mathcal{D}}}  \right\},\label{eq:conformal_threshold_sel}
  \end{align}
  then satisfies
  \begin{align*}
    \mathbb{E}_{(X,Y)\sim P_{XY}}\left[\ell\left(\Gamma_{\lambda^{*}}(X), Y\right)\right]\le \varepsilon.
  \end{align*}
\end{lemma}

Note that the assumption in \cref{eq:conformal_loss_assumption} is the same as stated in \cref{sec:metrics_problem_statement}. The term $(\gamma-\varepsilon)/N_{\mathcal{D}}$ in \cref{eq:conformal_threshold_sel} is a correction factor that accounts for the finite size of the calibration set, ensuring that the guarantee holds for unseen samples. As expected, a larger calibration set leads to a smaller correction term and thus a smaller prediction set. Furthermore, the result places no assumptions on the model $f(\cdot)$, other than it outputting a confidence score for each potential label $y\in\mathcal{Y}$. The threshold $\lambda^{*}$ can be computed by only evaluating the model on the calibration dataset samples.

To apply conformal risk control to our setting, we consider the entire encoder-decoder-inference pipeline as a single, composite black-box model $g_{l,k}$, defined as
\begin{equation}
  g_{l,k}(X) = f_k(d_l(e_l(X)))\label{eq:g_lk}
\end{equation}
for $1\le l\le L$ and $1\le k\le K$. Each of these composite models $g_{l,k}$ is then calibrated independently using the calibration dataset and the procedure outlined in \cref{lemma:conformalrisk} to obtain a specific threshold $\lambda_{l,k}$. However, the loss requirement in \cref{eq:problemdefstconformal} is conditioned on the event that the transmission succeeds within the frame, and thus $\alpha$ cannot be used directly in place of $\varepsilon$. To simplify the notation, let $\ell=\ell(\Gamma(X_t),Y_t)$, $T_{\le T}=T_{\text{tot},t}\le T$, and $T_{> T}=T_{\text{tot},t}> T$ define the loss and the events that the deadline is met and violated, respectively. By the law of total expectation we have
\begin{align*}
  \mathbb{E}[\ell\,|\,E_{\le T}] &= \frac{\mathbb{E}[\ell]-\mathbb{E}[\ell\,|\,T_{> T}]\Pr(T_{> T})}{1-\Pr(T_{>T})}\\
  &\le \frac{\mathbb{E}[\ell]}{1-\beta},
\end{align*}
where the inequality follows from the definition of $\beta$ and the fact that $\mathbb{E}[\ell\,|\,T_{> T}]\Pr(T_{> T})\ge 0$.
It follows that $\mathbb{E}[\ell]/(1-\beta)\le \alpha$ is a sufficient condition to satisfy constraint \cref{eq:problemdefstconformal}, which can be guaranteed through conformal risk control by choosing
\begin{align}
  \varepsilon = \alpha(1-\beta).\label{eq:corrected_risk_level_ub}
\end{align}

The combinatorial approach of performing conformal risk control on composite models might seem limiting in terms of the number of encoder/decoder and inference model combinations. However, it is generally necessary since the intermediate representations can vary significantly between different encoder/decoder pairs. Different encoder/decoders might use different compression ratios, or different feature extraction methods, leading to intermediate representations $Z$ with varying statistical properties and information content. Consequently, a calibration that works well for one encoder/decoder pair might be ineffective for another, necessitating individual calibration of each $g_{l,k}$. If the number of model combinations is prohibitive, alternative methods such as the learn-then-test framework~\cite{Angelopoulos2021LTT,zecchin2025allt} can be used to efficiently and jointly search for model combinations and prediction thresholds that satisfy the requirements. We leave such considerations for future work.

\section{Fixed Model Selection}\label{sec:fixed_selection}
In this section, we consider a fixed (offline) model selection scenario, wherein the encoder/decoder models and the edge inference model are selected a priori based only on the statistics of the channels, and these same models are executed in each frame. This scenario is relevant in a number of practical situations, such as when a priori model selection is required by the application, or in cases where the computational and latency overhead associated with online model selection is prohibitive. Throughout this section, we will focus on the restricted problem in~\eqref{eq:problemdef}.

With this setup, our objective is to construct a single composite model $g_{l,k}$ that satisfies constraints \eqref{eq:problemdefstconformal} and \eqref{eq:problemdefstoutage} while minimizing the expected size of the prediction set. Constraint \eqref{eq:problemdefstconformal} can be satisfied by any model combinations by employing conformal risk control to any composite model $g_{l,k}$ as described in \cref{sec:conformalrisk}, i.e., by constructing the output prediction set at the edge server as
\begin{align*}
  \Gamma(X) &=\left\{y\in\mathcal{Y}\, :\, [g_{l,k}(X)]_y \ge 1-\lambda_{l,k}\right\},
\end{align*}
where the threshold $\lambda_{l,k}$ is chosen by calibrating $g_{l,k}$ using $\mathcal{D}$ based on \cref{lemma:conformalrisk} with $\varepsilon = \alpha(1-\beta)$ as given by \cref{eq:corrected_risk_level_ub}.

On the other hand, the deadline violation constraint in \eqref{eq:problemdefstoutage} may not be satisfied by all models, since it depends on the instantaneous uplink and downlink communication rates $R_{\mathrm{ul},t}$ and $R_{\mathrm{dl},t}$, and on the size $D_{\mathrm{ul},l}$ of the encoded message in the uplink and $D_{\mathrm{dl},l,k}$ of the prediction set message in the downlink produced by the chosen model after employing conformal risk control. While the statistics of the instantaneous rates are assumed to be known, $D_{\mathrm{ul},l}$ and $D_{\mathrm{dl},l,k}$ depend on the chosen model combination and the distribution $P_{XY}$, and do not have known expressions. To overcome this, we propose a procedure, similar to conformal risk control, which uses only the empirical statistics obtained using the unlabeled calibration dataset $\mathcal{U}$, while carefully considering the effect of the finite number of samples.

The procedure relies on the following upper bound on the delay violation probability of a chosen model combination. The bound can be computed after the model thresholds $\lambda_{l,k}$ have been determined as outlined in \cref{sec:conformalrisk}.

\begin{proposition}[Delay violation bound]\label{prop:delay_violation}
  Consider a composite model $g_{l,k}$ as defined in \cref{eq:g_lk}. Let $\sigma_{\mathrm{ul},l}$ and $\sigma_{\mathrm{dl},l,k}$ be index permutations on $\{1,\ldots,N_{\mathcal{U}}\}$ that order the samples of the unlabeled calibration dataset $\mathcal{U}$ based on their uplink and downlink data sizes under the composite model $g_{l,k}$, respectively, in non-decreasing order:
  \begin{align*}
    &D_{\mathrm{ul},l}(X_{\sigma_{\mathrm{ul},l}(1)}^{(\mathcal{U})})\le\ldots\le D_{\mathrm{ul},l}(X_{\sigma_{\mathrm{ul},l}(N_{\mathcal{U}})}^{(\mathcal{U})}),\\
    &D_{\mathrm{dl},l,k}(X_{\sigma_{\mathrm{dl},l,k}(1)}^{(\mathcal{U})})\le\ldots\le D_{\mathrm{dl},l,k}(X_{\sigma_{\mathrm{dl},l,k}(N_{\mathcal{U}})}^{(\mathcal{U})}),
  \end{align*}
  i.e., $\sigma_{\mathrm{ul},l}(i)$ and $\sigma_{\mathrm{dl},l,k}(j)$ are the indices of the calibration samples with the $i$-th and $j$-th smallest uplink and downlink data sizes, respectively.
  The delay violation probability is then bounded as
  \begin{align*}
    &\Pr\left(T_{\text{tot},t}> T\,\middle|\,g_{l,k}\right)\nonumber\\
    &\qquad\le\min_{n,m\in\{1,\ldots,N_{\mathcal{U}}\}} 1- e^{\bar{\beta}_{\mathrm{cal}}(l,k,n,m)}\left(\frac{n+m}{N_{\mathcal{U}}+1} - 1\right),
  \end{align*}
  where
  \begin{align*}
    \textstyle\bar{\beta}_{\text{cal}}(l,\!k,\!n,\!m)\!\!=\!\!\left(\!\textsf{SNR}_{\mathrm{ul}}^{-1}\!\!+\!\textsf{SNR}_{\mathrm{dl}}^{-1}\!\right)\!\!\left(\!1\!-\!2^{\frac{\bar{D}_{\mathrm{ul},l}(n)+\bar{D}_{\mathrm{dl},l,k}(m)}{B\left(T-\tau_{\mathrm{ul},l}-\tau_{f_k}\right)}}\!\right),
  \end{align*}
  and
  \begin{align}
    \bar{D}_{\mathrm{ul},l}(n)&=D_{\mathrm{ul},l}\left(X_{\sigma_{\mathrm{ul},l}(n)}^{(\mathcal{U})}\right),\label{eq:def_bar_d_ul}\\
    \bar{D}_{\mathrm{dl},l,k}(m)&=D_{\mathrm{dl},l,k}\left(X_{\sigma_{\mathrm{dl},l,k}(m)}^{(\mathcal{U})}\right),\label{eq:def_bar_d_dl}
  \end{align}
  are the $n$-th and $m$-th order statistics of $\{D_{\mathrm{ul},l}(X_{n}^{(\mathcal{U})})\}_{n=1}^{N_{\mathcal{U}}}$ and $\{D_{\mathrm{dl},l,k}(X_{n}^{(\mathcal{U})})\}_{n=1}^{N_{\mathcal{U}}}$, respectively.
\end{proposition}
\begin{IEEEproof}
  See \cref{app:proof:prop:delay_violation}.
\end{IEEEproof}
Note that the delay violation probability bound in \cref{prop:delay_violation} is computed using the unlabeled dataset $\mathcal{U}$, which contains samples independent of the ones in $\mathcal{D}$ used for calibration, in order to ensure an unbiased estimate of the marginal distributions used in the bound.

\Cref{prop:delay_violation} enables us to determine whether a given model combination satisfies constraint \eqref{eq:problemdefstoutage} \emph{after} the prediction set threshold has been determined to meet constraint \eqref{eq:problemdefstconformal}. Leveraging this result along with conformal risk control from \cref{lemma:conformalrisk}, we propose the scheme listed in \cref{alg:fixedpolicy}, which relies on both the labeled and unlabeled calibration datasets $\mathcal{D}$ and $\mathcal{U}$. Specifically, for each composite model $g_{l,k}$, the edge server first applies \cref{lemma:conformalrisk} on the labeled dataset $\mathcal{D}$ to find a threshold $\lambda_{l,k}$ required to satisfy constraint \eqref{eq:problemdefstconformal} using the corrected risk level upper bound in \cref{eq:corrected_risk_level_ub} (\cref{alg:fixedpolicy:threshold}). Using the threshold $\lambda_{l,k}$, it then uses \cref{lemma:conditional_lb} with the unlabeled dataset $\mathcal{U}$ to compute the delay violation probability bound $\bar{P}_{l,k}$ (\cref{alg:fixedpolicy:p_bar}), and also to estimate the expected size of the prediction set $\bar{\Gamma}_{l,k}$ (\cref{alg:fixedpolicy:gamma_bar}).
The procedure then checks if model $g_{l,k}$ is better than the best one identified so far (\crefrange{alg:fixedpolicy:if_start}{alg:fixedpolicy:if_end}). Specifically, $g_{l,k}$ is better if it either (i) satisfies the required delay violation probability and has a smaller expected prediction set size, or (ii) if no models examined so far satisfy the delay violation requirement and $g_{l,k}$ exhibits a smaller delay violation probability.
This ensures that, even when no model combination meets the delay violation requirement, the procedure returns the model with the lowest estimated delay violation probability\footnote{Alternative strategies for handling the case where no model satisfies the requirement are straightforward to implement, but are beyond the scope of this discussion.}. Finally, the procedure returns the best model (\cref{alg:fixedpolicy:ret}).

The fixed model selection procedure in \cref{alg:fixedpolicy} does not depend on the instantaneous input $X_t$ or the instantaneous channel rates $R_{\mathrm{ul},t}$ and $R_{\mathrm{dl},t}$, the model selection procedure can be executed offline, and is thus suitable for resource-constrained environments.

\begin{algorithm}[t]
  \caption{Fixed Model Selection.}\label{alg:fixedpolicy}
  \begin{algorithmic}[1]\small
    \Function{FixedModelSelect}{$\{(e_l,d_l)\}_{l=1}^L, \{f_k\}_{k=1}^K, \mathcal{D}, \mathcal{U}, \alpha, \beta, T$}
      \State Initialize $g^* \gets \textsc{Null}$;  $\lambda^* \gets 0$; $\bar{\Gamma}^* \gets \infty$; $\bar{P}^* \gets \infty$.
      \For{$l=1,2,\ldots,L$}
        \For{$k=1,2,\ldots,K$}
          \State Define $\Gamma_{\lambda,l,k}(X) = \left\{y\in\mathcal{Y}\, :\, [g_{l,k}(X)]_y \ge 1-\lambda\right\}$.
          \State Compute the threshold $\lambda_{l,k}$ using \cref{lemma:conformalrisk} with $\mathcal{D}$
          \Statex\hspace*{\algorithmicindent}\hspace*{\algorithmicindent}\hspace{\algorithmicindent}\quad for $\Gamma_{\lambda,l,k}$ and $\varepsilon=\alpha(1-\beta)$.\label{alg:fixedpolicy:threshold}
          \State Compute $\bar{P}_{l,k}$ as the resulting delay violation
          \Statex\hspace*{\algorithmicindent}\hspace*{\algorithmicindent}\hspace{\algorithmicindent}\quad probability upper bound in \cref{prop:delay_violation} using $\mathcal{U}$.\label{alg:fixedpolicy:p_bar}
          \State $\bar{\Gamma}_{l,k} \gets \frac{1}{N_{\mathcal{U}}}\sum_{n=1}^{N_{\mathcal{U}}}|\Gamma_{\lambda_{l,k},l,k}(X_n^{(\mathcal{U})})|$.\label{alg:fixedpolicy:gamma_bar}
          \If{($\bar{P}_{l,k} \le \beta$ \emph{and} $\bar{\Gamma}_{l,k} < \bar{\Gamma}^*$)%
              \Statex\hspace*{\algorithmicindent}\hspace*{\algorithmicindent}\hspace{\algorithmicindent}\quad \emph{or} ($\bar{P}^{*} \ge \beta$ \emph{and} $\bar{P}_{l,k}< \bar{P}^{*}$)}\label{alg:fixedpolicy:if_start}
            \State $g^{*}\gets g_{l,k}$; $\lambda^* \gets \lambda_{l,k}$; $\bar{\Gamma}^* \gets \bar{\Gamma}_{l,k}$; $\bar{P}^{*}\gets\bar{P}_{l,k}$.\label{alg:fixedpolicy:if_end}
          \EndIf
        \EndFor
      \EndFor
      \State\Return $(g^{*}, \lambda^{*})$.\label{alg:fixedpolicy:ret}
    \EndFunction
  \end{algorithmic}
\end{algorithm}

\section{Dynamic Model Selection}\label{sec:dynamic_selection}
The fixed model selection presented in \cref{sec:fixed_selection} suffers from the fact that it must guarantee that the deadline is met over a wide range of channels, and thus the selected models are often overly conservative.
In this section, we extend the fixed model selection method to the case where the edge model can be selected dynamically \emph{after} the edge server has received the observation from the device. In general, this should allow for better performance since the edge model can be selected based on the actual time remaining before the deadline. However, guaranteeing the performance through conformal risk control while directly conditioning on the remaining time is non-trivial. This is because a short observation message will on average take shorter to transmit than a long message, which introduces a bias in the edge model input distribution, causing it to be different from the one used for calibration. For instance, having a long time available until the deadline is more likely when the message is short, but a short message may also be associated with a high inference uncertainty and consequently a large prediction set. Although it is possible to perform the calibration procedure conditioned on the time remaining before the deadline, e.g., in an ad-hoc/online fashion, the computational complexity associated with calibration makes such methods impractical. Instead, in this paper we propose to extend the model selection scheme in \cref{sec:fixed_selection} by conditioning only on the instantaneous rate of the uplink channel rather than the actual duration of the uplink transmission. Since the channel is independent of input observation, this does not introduce bias. Thus, while it ignores the specific time remaining until the deadline, it allows us to preserve the strong guarantees of the previous scheme without requiring online calibration.

We first present a dynamic model selection scheme for the problem in~\eqref{eq:problemdef}, and afterwards present a scheme for a slightly relaxed problem, which allows for more flexibility through truncation of the prediction set.

\subsection{Dynamic Model Selection for \eqref{eq:problemdef}}\label{sec:dynmodelproblemdef}

The proposed dynamic model selection procedure deviates from the fixed procedure in \cref{sec:fixed_selection} only in that the edge model is re-evaluated at the edge server after observing the uplink channel. Specifically, we first apply the fixed policy to select a composite model that meets the delay and loss constraints. From this composite model, only the encoder/decoder pair is used, while the edge server model is selected after observing the channel. This approach ensures that at least one edge model that meets the constraints exists (the one that was selected by the fixed policy). However, as argued above, it is likely that the instantaneous channel allows us to execute a larger and better model, and thus to obtain a smaller prediction set.

Compared to the fixed model selection procedure, the main component in the dynamic model selection is the following result, which is similar to \cref{prop:delay_violation} but conditioned on the instantaneous uplink channel.

\begin{proposition}[Conditional delay violation bound]\label{prop:delay_violation_dynamic}
  Consider a composite model $g_{l,k}$ as defined in \cref{eq:g_lk}. Let $\sigma_{\mathrm{ul},l}$ and $\sigma_{\mathrm{dl},l,k}$ be index permutations on $\{1,\ldots,N_{\mathcal{U}}\}$ that order the samples of the unlabeled calibration dataset $\mathcal{U}$ based on their uplink and downlink data sizes under the composite model $g_{l,k}$, respectively, in non-decreasing order:
  \begin{align*}
    &D_{\mathrm{ul},l}(X_{\sigma_{\mathrm{ul},l}(1)}^{(\mathcal{U})})\le\ldots\le D_{\mathrm{ul},l}(X_{\sigma_{\mathrm{ul},l}(N_{\mathcal{U}})}^{(\mathcal{U})}),\\
    &D_{\mathrm{dl},l,k}(X_{\sigma_{\mathrm{dl},l,k}(1)}^{(\mathcal{U})})\le\ldots\le D_{\mathrm{dl},l,k}(X_{\sigma_{\mathrm{dl},l,k}(N_{\mathcal{U}})}^{(\mathcal{U})}).
  \end{align*}
  Let $R_{\mathrm{ul},t}$ denote the instantaneous rate supported by the uplink channel.
  The delay violation probability conditioned in $R_{\mathrm{ul},t}$ is then bounded as
  \begin{align*}
    &\Pr\left(T_{\text{tot},t}> T\,\middle|\,R_{\mathrm{ul},t}, g_{l,k}\right)\nonumber\\
    &\qquad\le\min_{n,m\in\{1,\ldots,N_{\mathcal{U}}\}} 1- e^{\hat{\beta}_{\mathrm{cal}}(l,k,n,m)}\left(\frac{n+m}{N_{\mathcal{U}}+1} - 1\right),
  \end{align*}
  where
  \begin{align*}
    \textstyle\hat{\beta}_{\text{cal}}(l,k,n,m)\!=\!\textsf{SNR}_{\mathrm{dl}}^{-1}\left(\!1\!-\!2^{\frac{\bar{D}_{\mathrm{dl},l,k}(m)}{B\left(T-\tau_{\mathrm{ul},l}-\tau_{f_k}-\bar{D}_{\mathrm{ul},l}(n)/R_{\mathrm{ul},t}\right)}}\!\right),
  \end{align*}
  and
  \begin{align}
    \bar{D}_{\mathrm{ul},l}(n)&=D_{\mathrm{ul},t}\left(X_{\sigma_{\mathrm{ul},l}(n)}^{(\mathcal{U})}\right),\label{eq:def_bar_d_ul_dynamic}\\
    \bar{D}_{\mathrm{dl},l,k}(m)&=D_{\mathrm{dl},t}\left(X_{\sigma_{\mathrm{dl},l,k}(m)}^{(\mathcal{U})}\right),\label{eq:def_bar_d_dl_dynamic}
  \end{align}
  are the $n$-th and $m$-th order statistics of $\{D_{\mathrm{ul},l}(X_{n}^{(\mathcal{U})})\}_{n=1}^{N_{\mathcal{U}}}$ and $\{D_{\mathrm{dl},l,k}(X_{n}^{(\mathcal{U})})\}_{n=1}^{N_{\mathcal{U}}}$, respectively.
\end{proposition}
\begin{IEEEproof}
  See \cref{app:proof:prop:delay_violation_dynamic}.
\end{IEEEproof}

Note that, although the message is known to the edge server, \cref{prop:delay_violation_dynamic} is obtained by conditioning only on the channel rate $R_{\mathrm{ul},t}$, while treating the message size as a random variable. As mentioned, this is to avoid implicit bias caused by the fact that the message is likely to influence the size of the produced prediction sets.

Using \cref{prop:delay_violation_dynamic}, we propose the dynamic model selection algorithm listed in \cref{alg:dynpolicy}. The general procedure is similar to the fixed model selection scheme in \cref{alg:fixedpolicy} but differs in that it takes as input an encoder/decoder model $(e_l,d_l)$ selected using \cref{alg:fixedpolicy} and the instantaneous rate $R_{\mathrm{ul},t}$, and only outputs the edge model and the corresponding prediction set threshold $(f^{*},g^{*})$, rather than the composite model. Note that, while $R_{\mathrm{ul},t}$ is available only at inference time, the thresholds $\lambda_{l,k}$ in \cref{alg:dynpolicy:threshold} can be computed offline. On the other hand, the delay violation bound $\hat{P}_{l,k}$ in \cref{alg:dynpolicy:p_hat} and the proceeding steps depend on $R_{\mathrm{ul},t}$, and must be computed at inference time. Consequently, the dynamic policy comes at an additional computational cost compared to the fixed policy.

The proposed algorithm does not make use of the downlink channel rate, although, e.g., truncating the prediction set to guarantee successful transmission before the deadline would be a natural strategy. However, while such truncation would help satisfying the deadline violation requirement in \cref{eq:problemdefstoutage}, it is likely to lead to a violation of the loss requirement in \cref{eq:problemdefstconformal}. This is because the loss requirement is conditioned on the event of successful transmission, and truncation increases the successful transmission probability at the cost of an increase in the loss.
On the other hand, truncation of the prediction set can be performed under a slightly relaxed problem formulation, which we consider next.

\begin{algorithm}[t]
  \caption{Dynamic Model Selection}\label{alg:dynpolicy}
  \begin{algorithmic}[1]\small
    \Function{DynModelSelect}{$(e_l,d_l), \{f_k\}_{k=1}^K, \mathcal{D}, \mathcal{U}, \alpha, \beta, T, R_{\mathrm{ul},t}$}
      \State Initialize $f^* \gets \textsc{Null}$;  $\lambda^* \gets 0$; $\hat{\Gamma}^* \gets \infty$; $\hat{P}^* \gets \infty$.
        \For{$k=1,2,\ldots,K$}
          \State Define $\Gamma_{\lambda,l,k}(X) = \left\{y\in\mathcal{Y}\, :\, [g_{l,k}(X)]_y \ge 1- \lambda\right\}$.
          \State Compute the threshold $\lambda_{l,k}$ using \cref{lemma:conformalrisk} with $\mathcal{D}$
          \Statex\hspace*{\algorithmicindent}\hspace*{\algorithmicindent}\quad for $\Gamma_{\lambda,l,k}$ and $\varepsilon=\alpha(1-\beta)$.\label{alg:dynpolicy:threshold}
          \State Compute $\hat{P}_{l,k}$ as the resulting delay violation
          \Statex\hspace*{\algorithmicindent}\hspace{\algorithmicindent}\quad probability bound in \cref{prop:delay_violation_dynamic} using $\mathcal{U}$, $R_{\mathrm{ul},t}$.\label{alg:dynpolicy:p_hat}
          \State $\hat{\Gamma}_{l,k} \gets \frac{1}{N_{\mathcal{U}}}\sum_{n=1}^{N_{\mathcal{U}}}|\Gamma_{\lambda_{l,k},l,k}(X_n^{(\mathcal{U})})|$.\label{alg:dynpolicy:gamma_hat}
          \If{($\hat{P}_{l,k} \le \beta$ \emph{and} $\hat{\Gamma}_{l,k} < \hat{\Gamma}^*$)%
              \Statex\hspace*{\algorithmicindent}\hspace*{\algorithmicindent}\hspace{\algorithmicindent}\quad \emph{or} ($\hat{P}^{*} \ge \beta$ \emph{and} $\hat{P}_{l,k}< \hat{P}^{*}$)}\label{alg:dynpolicy:if_start}
            \State $f^{*}\gets f_{k}$; $\lambda^* \gets \lambda_{l,k}$; $\hat{\Gamma}^* \gets \hat{\Gamma}_{l,k}$; $\hat{P}^{*}\gets\hat{P}_{l,k}$.\label{alg:dynpolicy:if_end}
          \EndIf
        \EndFor
      \State\Return $(f^{*}, \lambda^{*})$.\label{alg:dynpolicy:ret}
    \EndFunction
  \end{algorithmic}
\end{algorithm}

\subsection{Dynamic Model Selection with Prediction Set Truncation}\label{sec:dynmodelrelaxedproblemdef}
As argued, the fact that the constraint in \eqref{eq:problemdefstconformal} is conditioned on successful transmission prevents us from truncating the prediction set based on the instantaneous rate $R_{\mathrm{ul},t}$ and the remaining time until the deadline, although doing so would increase the probability of meeting the deadline.
In this section we consider a slightly relaxed variant of the problem in~\eqref{eq:problemdef}, under which such truncation is possible. Specifically, we consider the problem
\begin{subequations}\label{eq:relaxedproblemdef}
  \begin{align}
    \minimize \quad & \mathbb{E}\left[|\Gamma(X_t)|\,\mid\,T_{\text{tot},t}\le T\right],\label{eq:relaxedproblemdefobj}\\
    \textrm{s.t.} \quad & \mathbb{E}\left[\ell'\left(\Gamma(X_t),Y_t\right)\right] \le \alpha',\label{eq:relaxedproblemdefstconformal}
  \end{align}
\end{subequations}
where
\begin{equation}\label{eq:ellprime}
    \ell'\left(\Gamma(X), Y\right) =
    \begin{cases}
      \ell\left(\Gamma(X), Y\right), & T_{\text{tot},t}\le T,\\
      \gamma, & \text{otherwise}.
    \end{cases}
  \end{equation}
  This is a relaxed problem since any solution to~\eqref{eq:problemdef} satisfies
  \begin{align}
    \mathbb{E}\left[\ell'\left(\Gamma(X_t), Y_t\right)\right] \le (1-\beta)\alpha + \beta\gamma.\label{eq:relaxed_alpha_equivalent}
  \end{align}
  However, a solution to~\eqref{eq:relaxedproblemdef} does generally not satisfy the constraints in \eqref{eq:problemdef}. Note that the constraint in \eqref{eq:relaxedproblemdefstconformal} has a natural interpretation when the loss is an indicator function, such as the 0--1 loss, where it bounds the probability of receiving a prediction set with zero loss before the deadline.

To see why truncation is possible while satisfying the constraint in~\eqref{eq:relaxedproblemdefstconformal}, consider a composite model obtained using the dynamic model selection procedure in \cref{alg:dynpolicy}, which satisfies the constraint in \eqref{eq:relaxedproblemdefstconformal} with $\alpha'=(1-\beta)\alpha + \beta\gamma$. Since the constraint in \eqref{eq:relaxedproblemdefstconformal} is not conditioned on successful transmission before the deadline, but instead assigns the maximum loss $\gamma$ to failed transmissions, truncating prediction sets that would otherwise fail can only reduce the expected loss. 
Note, however, that depending on the specific distribution of the prediction set sizes truncation may result in a larger conditional expected prediction set size defined as the objective in \eqref{eq:relaxedproblemdefobj}. Therefore, the advantage of truncation ultimately depends on the specific application.

To keep the presentation consistent with solutions to the problem in~\eqref{eq:problemdef} and to ease the comparison, we will assume that $\alpha'=(1-\beta)\alpha + \beta\gamma$ for some specified values of $\alpha$ and $\beta$. In this case, truncation can be implemented on top of the dynamic model selection scheme in \cref{alg:dynpolicy}. Specifically, we truncate the prediction set constructed by the edge server model selected by \cref{alg:dynpolicy} to at most
\begin{equation*}
  \tilde{\Gamma}_t=\max\left(1, \left\lfloor\frac{R_{\mathrm{dl},t}(T-\tau_{\mathrm{ul},l}-\tau_{f_k}-T_{\mathrm{ul},t})}{D_{\mathrm{lbl}}}\right\rfloor\right),
\end{equation*}
so that the resulting prediction set is given as
\begin{equation*}
\begin{split}
  &\Gamma(X_t)\\
  &\quad=\!\left\{y\in \mathcal{Y} : [g_{l,k}(X_t)]_y \ge 1-\lambda_{l,k},y\in\mathrm{top}_{\tilde{\Gamma}_t}(g_{l,k}(X_t))\right\},
\end{split}
\end{equation*}
where $\mathrm{top}_{\tilde{\Gamma}_t}(g_{l,k}(X_t))$ is the set of $\tilde{\Gamma}_t$ labels with the largest scores.
As discussed, whether truncation improves the resulting conditional expected size of the prediction set depends on the specific problem, but the expected (modified) loss $\mathbb{E}\left[\ell'\left(\Gamma(X_t), Y_t\right)\right]$ will be less than or equal to that achieved by applying the procedure without truncation.

\section{Numerical Results}\label{sec:numres}
We demonstrate the proposed framework through numerical results. We first outline the setup and baselines. We then present results under the joint loss and deadline guarantees in~\eqref{eq:problemdef}, followed by results for under the relaxed loss with prediction set truncation (the problem in~\eqref{eq:relaxedproblemdef}).

\subsection{Experimental Setup and Baselines}
\subsubsection{Experimental Setup}
We evaluate the proposed framework on an image classification task with the ImageNet 2012 dataset~\cite{imagenet}, so that the sensor input $X_t$ is an image and $Y_t$ is the ground-truth image category, representing one of the 1000 ImageNet classes.
We consider the models listed in \cref{tab:models}. Specifically, the encoder/decoder models are implemented as the WebP~\cite{webp} image compression algorithm under various quality settings using the Python Pillow package. The WebP algorithm generally offers better compression than traditional formats like JPEG, and has a widely tunable trade-off between quality, size, and computation time. The intermediate representation $Z_t$ is thus the uncompressed image. The edge inference models are realized using EfficientNetV2~\cite{pmlr-v139-tan21a} classifier models, which are available in small, medium, and large variants. We use the model implementation from the PyTorch~\cite{pytorch} framework, pretrained on the ImageNet dataset. The computation times for the encoders/decoders selected for illustrative purposes, but based on rough estimates obtained using the ImageNet dataset and thus represent realistic numbers, while the computation times for the classifier models are the ones reported in~\cite{pmlr-v139-tan21a}.

We consider the 0--1 missed detection loss
\begin{equation*}
  \ell(\Gamma(X),Y)=\mathbbm{1}[Y\notin \Gamma(X)],
\end{equation*}
i.e., $\gamma=1$. The deadline is $T=150~\text{ms}$, and the expected loss and deadline violation probability requirements are set to $\alpha=0.01$ and $\beta=0.01$, respectively. Each predicted label $y\in\mathcal{Y}$ is assumed to occupy $D_{\mathrm{lbl}}=64~\mathrm{bits}$, and the bandwidth is $B=30~\text{MHz}$.
The ImageNet validation dataset is randomly split into three disjoint sets for calibration ($N_{\mathcal{D}}=10000$ labeled and $N_{\mathcal{U}}=10000$ unlabeled) and evaluation ($30000$ labeled).

\begingroup
\setlength{\tabcolsep}{2pt}
\begin{table}[tb]
\centering\footnotesize
\caption{Models used in the numerical results}\label{tab:models}\vspace{-1em}
\begin{tabular}{lll}
  \toprule
  Encoder/decoder, $(e_l,d_l)$ & WebP quality setting & Computation time, $\tau_{\mathrm{ul},l}$ \\\midrule
  WebP-0, $(e_1,d_1)$       & $0$                                    & $10.0$ ms                           \\
  WebP-20, $(e_2,d_2)$       & $20$                                    & $12.5$ ms                           \\
  WebP-50, $(e_3,d_3)$       & $50$                                    & $15.0$ ms                           \\
  WebP-80, $(e_4,d_4)$       & $80$                                    & $17.5$ ms                           \\\midrule
Classifier model, $f_k$ & Num.\ parameters & Computation time, $\tau_{\mathrm{dl},k}$ \\\midrule
EfficientNetV2-S, $f_1$       & $22$M                                    & $24.0$ ms                           \\
EfficientNetV2-M, $f_2$      & $54$M                                   & $57.0$ ms                           \\
EfficientNetV2-L, $f_3$     & $120$M                                   & $98.0$ ms                          \\\bottomrule
\end{tabular}
\end{table}
\endgroup

\subsubsection{Baselines}
We compare our proposed model selection framework to a small and a large fixed model execution policy, which always execute the same model regardless of the SNR. The small baseline model comprises the WebP-0 encoder/decoder and the EfficientNetV2-S classifier, i.e., $g_{1,1}$, while the large baseline model is defined by WebP-80 and EfficientNetV2-L, i.e., $g_{4,3}$. For each model, we consider both a Top-$20$ aggregation function, which outputs a prediction set containing the $20$ labels with the largest confidence scores, and the calibrated threshold-based conformal aggregation function presented in \cref{sec:conformalrisk}.

\begin{figure}
  \centering
  \subfloat[Average loss vs.\ SNR]{\includegraphics{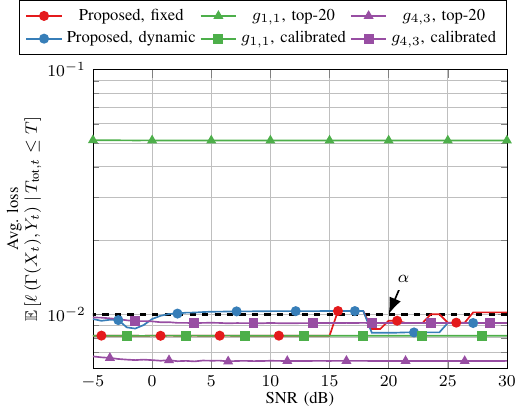}\label{fig:res_joint_avg_loss}}\\[-0.1em]
  \subfloat[Deadline violation probability vs.\ SNR]{\includegraphics{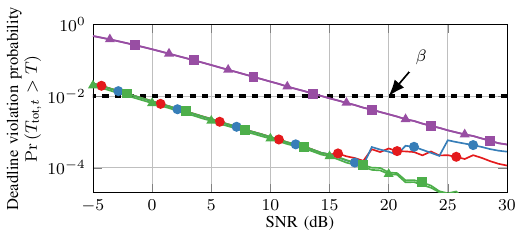}\label{fig:res_joint_deadline_violation}}\\[-0.1em]
  \subfloat[Average prediction set size vs. SNR]{\includegraphics{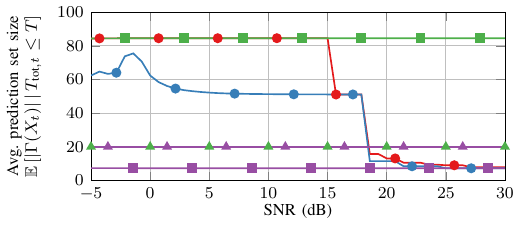}\label{fig:res_joint_pred_size}}
  \caption{Model performance of the proposed schemes and baselines vs. SNR in terms of \protect\subref{fig:res_joint_avg_loss} average loss, \protect\subref{fig:res_joint_deadline_violation} deadline violation probability, and \protect\subref{fig:res_joint_pred_size} prediction set size.}\label{fig:res_loss_deadline_size}
\end{figure}

\subsection{Joint Loss and Deadline Guarantees}\label{sec:res_joint_loss_deadline}
In this section, we study the proposed fixed and dynamic model selection schemes for the problem in~\eqref{eq:problemdef}. \Cref{fig:res_loss_deadline_size} compares the proposed schemes to the baselines in terms of loss, deadline violation probability, and prediction set size. \Cref{fig:res_joint_avg_loss} shows that the loss of the proposed schemes (solid red and blue) is very close to the loss requirement of $\alpha=0.01$ (indicated by the dashed black line), thereby achieving the desired loss. Similarly, the calibrated baseline models (green and purple, square marker) achieve the desired loss, as expected by the calibration procedure. The baselines that output the top-$20$ prediction set, only the large model combination $g_{4,3}$ meets the desired loss, while the small model $g_{1,1}$ has a high loss. This demonstrates both the trade-off between representation and model complexity and inference quality, and also highlights the importance of calibrating the individual model combinations. The minor deviations from the loss requirement $\alpha$ seen for the proposed schemes at certain SNRs are attributed to finite sample effects in the evaluation set, which diminish with larger evaluation datasets.

While the $g_{4,3}$ baseline models meet the loss requirement, \cref{fig:res_joint_deadline_violation} reveals that it fails to meet the deadline violation requirement of $\beta=0.01$ over a large range of SNRs. On the other hand, all other models meet the requirement except at very low SNRs. In this regime, none of the available model combinations in \cref{tab:models} meet the requirement, and thus it is not possible to satisfy the requirement. Nevertheless, it can be seen that the proposed schemes achieve the smallest possible deadline violation probability as desired. At high SNRs, the deadline violation probability of the proposed schemes deviate from the baselines. This is because a high SNR allows the schemes to select a better model combination while still satisfying the deadline requirement. In this high SNR regime, the deadline violation probability is generally quite far from the required $\beta$. This is because the bounds in \cref{prop:delay_violation,prop:delay_violation_dynamic} are derived under the assumption that the uplink and downlink transmission delays are correlated, which is a conservative assumption.

The significance of selecting the model based on the SNR is reflected in the average prediction set size (\cref{fig:res_joint_pred_size}), where the prediction set size of the proposed schemes decreases as the SNR increases. In particular, at low SNRs a small model is required to satisfy the requirement, resulting in a large prediction set. However, as the SNR increases, the schemes gradually select larger models, resulting in smaller and more informative prediction sets. Conversely, the baselines fail to adapt to the SNR, resulting in a constant prediction set size across SNRs. \Cref{fig:res_joint_pred_size} also shows the benefit of the dynamic model selection scheme, which at low SNRs has a significantly smaller prediction set size compared to the fixed model selection scheme.

\begin{figure}
  \centering
  \subfloat[Fixed model selection]{\includegraphics{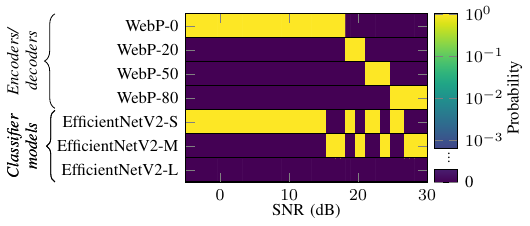}\label{fig:res_model_selection_fixed}}\\[-0.2em]
  \subfloat[Dynamic model selection]{\includegraphics{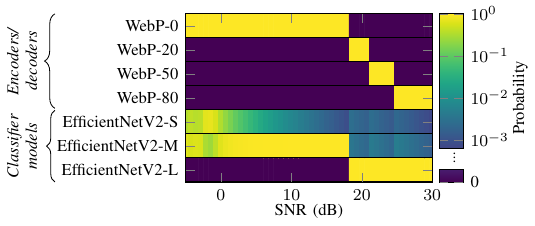}\label{fig:res_model_selection_dynamic}}
  \caption{Distribution of the models selected by the proposed model selection schemes vs. SNR. The color indicates the probability.}\label{fig:res_model_sel}
\end{figure}

The distributions of the model combinations selected by the proposed fixed and dynamic model selection schemes are illustrated in \cref{fig:res_model_selection_fixed,fig:res_model_selection_dynamic}, respectively. As the SNR increases, the fixed model selection algorithm \cref{fig:res_model_selection_fixed} gradually selects higher quality encoder/decoder model, while the classifier models alternate between EfficientNetV2-S and EfficientNetV2-M. The dynamic model selection scheme (\cref{fig:res_model_selection_dynamic}) has a similar behavior, but the probabilistic nature of the scheme makes the transition more smooth. Note that the dynamic model selects the encoder/decoder model in the same way as the fixed scheme, and thus executes the same encoder/decoder as the fixed scheme at each SNR. However, contrary to the fixed scheme, the dynamic scheme frequently executes EfficientNetV2-M at low SNR and EfficientNetV2-L at high SNR. Thus, the dynamic nature of the dynamic model selection results in a much less conservative model selection compared to the fixed model selection scheme. Note that EfficientNetV2-L is never chosen together with the WebP-0 encoder/decoder model. This because EfficientNetV2-L performs poorly on low-quality images, and confirms the proposed framework's ability to handle intricate model interactions. \Cref{fig:res_model_selection_fixed} also reveals interesting behavior at low SNRs, where the most likely inference model alternates between EfficientNetV2-S and EfficientNetV2-M. This can be explained as follows. At low SNRs, the communication dominates the total delay, and thus EfficientNetV2-M is selected more frequently as it generally outputs a smaller prediction set than EfficientNetV2-S. As the SNR increases, the communication delay becomes less dominant and EfficientNetV2-S, which is faster to execute, is preferred. Finally, the objective in~\eqref{eq:problemdefobj} of minimizing the expected prediction set size becomes the dominant factor for model selection, which again gives preference to EfficientNetV2-M.

\subsection{Relaxed Loss and Prediction Set Truncation}
\begin{figure}
  \centering
  \subfloat[Average relaxed loss]{\includegraphics{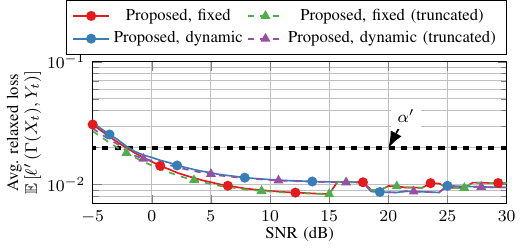}\label{fig:res_relaxed_loss}}\\[-0.1em]
  \subfloat[Average prediction set size]{\includegraphics{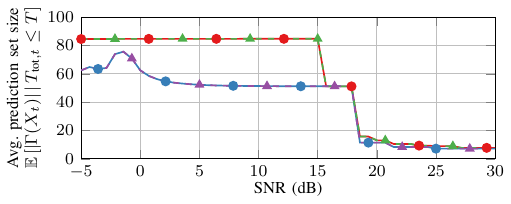}\label{fig:res_relaxed_avg_size}}
  \caption{Performance of the proposed schemes and baselines vs. SNR under the relaxed loss.}\label{fig:res_relaxed_loss_size}
\end{figure}

In this section, we evaluate the model selection procedures under the relaxed loss $\ell'$ defined in \eqref{eq:ellprime}, which allows for dynamic truncation of the prediction set depending on the instantaneous channel. For simplicity, we consider $\alpha'=(1-\beta)\alpha + \beta\gamma$, where $\alpha=0.01$ and $\beta=0.01$ as before (i.e., $\alpha'=0.0199$). \Cref{fig:res_relaxed_loss,fig:res_relaxed_avg_size} show the resulting relaxed loss and average prediction set size, respectively. Note that schemes without truncation are equivalent to the ones evaluated in \cref{sec:res_joint_loss_deadline}, but evaluated under the relaxed loss $\ell$. As can be seen from the figure, the schemes with truncation provide a slightly smaller loss at low SNRs, but otherwise performs similar to the schemes without truncation. This is because the transmission of the prediction set only constitutes a small fraction of the total delay budget, and thus has limited impact on the deadline violation probability.

\section{Conclusion}\label{sec:conclusion}
This paper presented a framework for black-box real-time edge AI under strict loss and deadline requirements. We assumed that the sensor and edge server have access to an ensemble of black-box encoder/decoder and inference models with various complexities and computation times. Leveraging conformal risk control and non-parametric statistics, we developed two model selection schemes that aim to maximize the informativeness of the predictions for given loss and deadline violation probability requirements. The first scheme executes the same model for a given SNR, while the second scheme dynamically adapts the edge model based on the instantaneous channel. Through numerical results of an image classification scenario, we demonstrated that the proposed framework meets the loss and deadline requirements while minimizing the average size of the prediction sets. Overall, this work establishes a new and general approach toward guaranteeing the end-to-end reliability and latency in integrated communication and AI scenarios, laying the foundation for reliable real-time edge AI in 6G.

\appendices
\crefalias{section}{appendix}

\section{Proof of \texorpdfstring{\cref{prop:delay_violation}}{Proposition~\ref{prop:delay_violation}}}\label{app:proof:prop:delay_violation}
We will prove \cref{prop:delay_violation} by first establishing a lower bound on the conditional probability $\Pr(T_{\mathrm{tot},t}\le T \,|\, D_{\mathrm{ul},t},D_{\mathrm{dl},t},g_{l,k})$, and then use it to bound the desired marginal probability $\Pr(T_{\mathrm{tot},t}> T \,|\, g_{l,k})$.
We first present the bound on $\Pr(T_{\mathrm{tot},t}\le T \,|\, D_{\mathrm{ul},t},D_{\mathrm{dl},t},g_{l,k})$.
\begin{lemma}\label{lemma:conditional_lb}
  For a composite model $g_{l,k}$ satisfying $\tau_{\mathrm{ul},l}+\tau_{f_k}\le T$, the conditional probability of satisfying the deadline given $D_{\mathrm{ul},t}$ and $D_{\mathrm{dl},t}$ is lower bounded as
  \begin{align*}
    &\Pr(T_{\mathrm{tot},t}\le T \,|\, D_{\mathrm{ul},t},D_{\mathrm{dl},t},g_{l,k})\nonumber\\
    &\quad\ge \exp\left(\left(\textsf{SNR}_{\mathrm{ul}}^{-1}+\textsf{SNR}_{\mathrm{dl}}^{-1}\right)\left(1-2^\frac{D_{\mathrm{ul},t}+D_{\mathrm{dl},t}}{B(T-\tau_{\mathrm{ul},l}-\tau_{f_k})}\right)\right).
  \end{align*}
\end{lemma}
\begin{IEEEproof}
  See \cref{app:proof:lemma:conditional_lb}.
\end{IEEEproof}

We now proceed to bound $\Pr(T_{\mathrm{tot},t}> T \,|\, g_{l,k})$.
From the law of total probability,
\begin{align*}
  &\Pr(T_{\mathrm{tot},t} > T\,|\,g_{l,k})\nonumber\\
  &\quad=1-\int_{0}^{\infty}\int_{0}^{\infty} \Pr(T_{\mathrm{tot},t} \le T \,|\, D_{\mathrm{ul},t}=\xi, D_{\mathrm{dl},t}=\psi,g_{l,k})\nonumber\\
  &\quad\qquad\quad\times p(D_{\mathrm{ul},t}=\xi,D_{\mathrm{dl},t}=\psi \,|\, g_{l,k})\,\mathrm{d}\xi\,\mathrm{d}\psi.
\end{align*}
Since the CDF is non-negative, restricting the domain of integration yields a lower bound on the integral. Thus, for any $D_{\mathrm{ul},t}'$ and $D_{\mathrm{dl},t}'$,
\begin{align}
  &\Pr(T_{\mathrm{tot},t} > T \,|\, g_{l,k})\nonumber\\
  &\quad\le\! 1\!-\! \int_{0}^{D_{\mathrm{ul},t}'}\!\int_{0}^{D_{\mathrm{dl},t}'}\!\Pr(T_{\mathrm{tot},t}\! \le\! T| D_{\mathrm{ul},t}\!=\!\xi,D_{\mathrm{dl},t}\!=\!\psi,g_{l,k})\nonumber\\
  &\quad\qquad\quad\times p(D_{\mathrm{ul},t}=\xi,D_{\mathrm{dl},t}=\psi \,|\, g_{l,k})\,\mathrm{d}\xi\,\mathrm{d}\psi\nonumber \\
  &\quad\overset{(a)}{\le} 1- \Pr(T_{\mathrm{tot},t} \le T \,|\, D_{\mathrm{ul},t}=D_{\mathrm{ul},t}',D_{\mathrm{dl},t}=D_{\mathrm{dl},t}',g_{l,k})\nonumber\\
  &\quad\qquad\quad\times \int_{0}^{D_{\mathrm{ul},t}'}\!\int_{0}^{D_{\mathrm{dl},t}'}\!p(D_{\mathrm{ul},t}\!=\!\xi,D_{\mathrm{dl},t}\!=\!\psi | g_{l,k})\,\mathrm{d}\xi\,\mathrm{d}\psi\nonumber\\
  &\quad\overset{(b)}{=} 1-\Pr(T_{\mathrm{tot},t} \le T \,|\, D_{\mathrm{ul},t}=D_{\mathrm{ul},t}',D_{\mathrm{dl},t}=D_{\mathrm{dl},t}',g_{l,k})\nonumber\\
  &\quad\qquad\quad\times \Pr(D_{\mathrm{ul},t}\le D_{\mathrm{ul},t}',D_{\mathrm{dl},t}\le D_{\mathrm{dl},t}' \,|\, g_{l,k}).\label{eq:lemma_marginal_bound_tmp0}
\end{align}
Here, inequality $(a)$ follows from the fact that $\Pr(T_{\mathrm{tot},t} \le T \,|\, D_{\mathrm{ul},t},D_{\mathrm{dl},t},g_{l,k})$ is nonincreasing in $D_{\mathrm{ul},t}$ and $D_{\mathrm{dl},t}$, and equality $(b)$ is obtained by noting that the integral evaluates to the joint CDF.

Let $L(D_{\mathrm{ul},t},D_{\mathrm{dl},t},l,k)$ denote the lower bound given in \cref{lemma:conditional_lb}, i.e.,
\begin{align*}
  &L(D_{\mathrm{ul},t},D_{\mathrm{dl},t},l,k)\nonumber\\
  &\quad=\exp\left(\left(\textsf{SNR}_{\mathrm{ul}}^{-1}+\textsf{SNR}_{\mathrm{dl}}^{-1}\right)\left(1-2^\frac{D_{\mathrm{ul},t}+D_{\mathrm{dl},t}}{B(T-\tau_{\mathrm{ul},l}-\tau_{f_k})}\right)\right).
\end{align*}
Substituting into \eqref{eq:lemma_marginal_bound_tmp0} gives us the bound
\begin{align}
  &\Pr(T_{\mathrm{tot},t} > T \,|\, g_{l,k})\nonumber\\
  &\quad\le 1- L(D_{\mathrm{ul},t}', D_{\mathrm{dl},t}',l,k)\nonumber\\
  &\quad\qquad\quad\times\Pr(D_{\mathrm{ul},t}\le D_{\mathrm{ul},t}',D_{\mathrm{dl},t}\le D_{\mathrm{dl},t}' \,|\, g_{l,k}).\label{eq:lemma_marginal_bound_tmp1}
\end{align}

The probability $\Pr(D_{\mathrm{ul},t}\le D_{\mathrm{ul},t}',D_{\mathrm{dl},t}\le D_{\mathrm{dl},t}' \,|\, g_{l,k})$ is unknown as it depends on the black-box encoder/decoder models $(e_l,d_l)$ and the edge model $f_k$ and the unknown distribution $P_{X}$. Furthermore, since $D_{\mathrm{ul},t}$ and $D_{\mathrm{dl},t}$ depend on the same input, they are in general not independent.
Instead, we proceed to derive a lower bound on $\Pr(D_{\mathrm{ul},t}\le D_{\mathrm{ul},t}',D_{\mathrm{dl},t}\le D_{\mathrm{dl},t}' \,|\, g_{l,k})$ using the unlabeled calibration dataset $\mathcal{U}$. The bound is presented in the following lemma.
\begin{lemma}\label{lemma:joint_size_lb}
  Let $\bar{D}_{\mathrm{ul},l}(n)$ and $\bar{D}_{\mathrm{dl},l,k}(m)$ be defined as in \cref{eq:def_bar_d_ul,eq:def_bar_d_dl}. Then, for any $n,m\in\{1,\ldots,N_{\mathcal{U}}\}$,
  \begin{align*}
    &\Pr(D_{\mathrm{ul},t}\!\le\! \bar{D}_{\mathrm{ul},l}(n),D_{\mathrm{dl},t}\!\le\! \bar{D}_{\mathrm{dl},l,k}(m) \,|\, g_{l,k})\!\ge\! \frac{n+m}{N_{\mathcal{U}}+1}-1,
  \end{align*}
  where the probability is over $X\sim P_{X}$.
\end{lemma}
\begin{IEEEproof}
  See \cref{app:proof:lemma:joint_size_lb}.
\end{IEEEproof}

Combining \cref{lemma:joint_size_lb} with \cref{eq:lemma_marginal_bound_tmp1}, and setting $D_{\mathrm{ul},t}'=\bar{D}_{\mathrm{ul},l}(n)$ and $D_{\mathrm{dl},t}'=\bar{D}_{\mathrm{dl},l,k}(m)$ yields
\begin{align*}
  &\Pr(T_{\mathrm{tot},t} > T \,|\, g_{l,k})\nonumber\\
  &\quad\textstyle\le 1- L\left(\bar{D}_{\mathrm{ul},l}(n), \bar{D}_{\mathrm{dl},l,k}(m),l,k\right)\left(\frac{n}{N_{\mathcal{U}}+1}+\frac{m}{N_{\mathcal{U}}+1}-1\right)\\
  &\quad\textstyle=1- L\left(\bar{D}_{\mathrm{ul},l}(n), \bar{D}_{\mathrm{dl},l,k}(m),l,k\right)
  \left(\frac{n+m}{N_{\mathcal{U}}+1}-1\right).      
\end{align*}
for any $n,m\in\{1,\ldots,N_{\mathcal{U}}\}$. To complete the proof, define
\begin{align*}
  \textstyle\bar{\beta}_{\mathrm{cal}}(l,k,n,m)&=\ln L\left(\bar{D}_{\mathrm{ul},l}(n), \bar{D}_{\mathrm{dl},l,k}(m),l,k\right)\\
  &\textstyle=\!\!\left(\!\textsf{SNR}_{\mathrm{ul}}^{-1}\!\!+\!\textsf{SNR}_{\mathrm{dl}}^{-1}\!\right)\!\left(\!1\!\!-\!\!2^\frac{\bar{D}_{\mathrm{ul},l}(n)+\bar{D}_{\mathrm{dl},l,k}(m)}{B(T-\tau_{\mathrm{ul},l}-\tau_{f_k})}\!\right),
\end{align*}
and choose $n$ and $m$ such that the bound is minimized.

\section{Proof of \texorpdfstring{\cref{lemma:conditional_lb}}{Lemma~\ref{lemma:conditional_lb}}}\label{app:proof:lemma:conditional_lb}
For any $0\le \phi\le T$ we have
\begin{align*}
  &\Pr(T_{\mathrm{tot},t}\le T\,|\,D_{\mathrm{ul},t},D_{\mathrm{dl},t},g_{l,k})\nonumber\\
  &\quad=\Pr(T_{\mathrm{ul},t}+T_{\mathrm{dl},t}\le T\,|\,D_{\mathrm{ul},t},D_{\mathrm{dl},t},g_{l,k})\\
  &\quad\ge \Pr(T_{\mathrm{ul},t}\le T-\phi, T_{\mathrm{dl},t}\le \phi\, |\,D_{\mathrm{ul},t},D_{\mathrm{dl},t},g_{l,k})\\
  &\quad= \Pr(T_{\mathrm{ul},t}\le T-\phi \,|\, D_{\mathrm{ul},t},g_{l,k})\Pr(T_{\mathrm{dl},t}\le \phi\,|\,D_{\mathrm{dl},t},g_{l,k}),
\end{align*}
where the second inequality follows from the fact that $T_{\mathrm{ul},t}$ and $T_{\mathrm{dl},t}$ are conditionally independent given $D_{\mathrm{ul},t},D_{\mathrm{dl},t},g_{l,k}$ (since $h_{\mathrm{ul},t}$ and $h_{\mathrm{dl},t}$ are independent), and that $T_{\mathrm{ul},t}$ is independent of $D_{\mathrm{dl},t}$, while $T_{\mathrm{dl},t}$ is independent of $D_{\mathrm{dl},t}$.
Expanding the terms first using \cref{eq:T_ul,eq:T_dl} and then using \cref{eq:capacity_ul,eq:capacity_dl} yields
\begin{align}
  &\Pr(T_{\mathrm{ul},t}\le T-\phi \,|\, D_{\mathrm{ul},t},g_{l,k})\Pr(T_{\mathrm{dl},t}\le \phi\,|\,D_{\mathrm{dl},t},g_{l,k})\nonumber\\
  &\quad\textstyle= \Pr\left(R_{\mathrm{ul},t}\!\ge\! \frac{D_{\mathrm{ul},t}}{T-\phi-\tau_{\mathrm{ul},l}} \,\middle|\, D_{\mathrm{ul},t}\right)\!\Pr\left(R_{\mathrm{dl},t}\!\ge\!\frac{D_{\mathrm{dl},t}}{\phi-\tau_{f_{k}}} \,\middle|\, D_{\mathrm{dl},t}\right)\nonumber\\
  &\quad\textstyle\overset{(a)}{=} \Pr\left(|h_{\mathrm{ul},t}|^2\ge \frac{2^{\frac{D_{\mathrm{ul},t}}{B\left(T-\phi-\tau_{\mathrm{ul},l}\right)}}-1}{\textsf{SNR}_{\mathrm{ul}}} \,\middle|\, D_{\mathrm{ul},t}\right)\nonumber\\
  &\qquad\quad\textstyle\times\Pr\left(|h_{\mathrm{dl},t}|^2\ge \frac{2^{\frac{D_{\mathrm{dl},t}}{B\left(\phi-\tau_{f_{k}}\right)}}-1}{\textsf{SNR}_{\mathrm{dl}}} \,\middle|\, D_{\mathrm{dl},t}\right)\nonumber\\
  &\quad\textstyle= \exp\left(\frac{1-2^{\frac{D_{\mathrm{ul},t}}{B(T-\phi-\tau_{\mathrm{ul},l})}}}{\textsf{SNR}_{\mathrm{ul}}}\right)
  \exp\left(\frac{1-2^{\frac{D_{\mathrm{dl},t}}{B(\phi-\tau_{f_k})}}}{\textsf{SNR}_{\mathrm{dl}}}\right)\nonumber\\
  &\quad=\textstyle \exp\left(\frac{1-2^{\frac{D_{\mathrm{ul},t}}{B(T-\phi-\tau_{\mathrm{ul},l})}}}{\textsf{SNR}_{\mathrm{ul}}}+\frac{1-2^{\frac{D_{\mathrm{dl},t}}{B(\phi-\tau_{f_k})}}}{\textsf{SNR}_{\mathrm{dl}}}\right),\label{eq:T_ul_bound_tmp}
\end{align}
where equality $(a)$ comes from the fact that $|h_{\text{ul},t}|^2$ and $|h_{\text{dl},t}|^2$ are exponentially distributed following the assumption of Rayleigh fading.

The best bound is obtained by maximizing $\phi\in [0,T]$. The derivative of the logarithm of \eqref{eq:T_ul_bound_tmp} is
\begin{align}
  &\frac{\mathrm{d}}{\mathrm{d}\phi}\left(\frac{1-2^{\frac{D_{\mathrm{ul},t}}{B(T-\phi-\tau_{\mathrm{ul},l})}}}{\textsf{SNR}_{\mathrm{ul}}}+\frac{1-2^{\frac{D_{\mathrm{dl},t}}{B(\phi-\tau_{f_k})}}}{\textsf{SNR}_{\mathrm{dl}}}\right)\nonumber\\
  &\quad\textstyle=-\textsf{SNR}_{\mathrm{ul}}^{-1}\frac{\mathrm{d}}{\mathrm{d}\phi}\left(2^{\frac{D_{\mathrm{ul},t}}{B(T-\phi-\tau_{\mathrm{ul},l})}}\right)\!-\! \textsf{SNR}_{\mathrm{dl}}^{-1}\frac{\mathrm{d}}{\mathrm{d}\phi}\left(2^{\frac{D_{\mathrm{dl},t}}{B(\phi-\tau_{f_k})}}\right)\nonumber\\
  &\quad\textstyle=-\textsf{SNR}_{\mathrm{ul}}^{-1}D_{\mathrm{ul},t}\log_2(D_{\mathrm{ul},t})\left(\frac{2^{\frac{D_{\mathrm{ul},t}}{B(T-\phi-\tau_{\mathrm{ul},l})}}}{B(T-\phi-\tau_{\mathrm{ul},l})^2}\right)\nonumber\\
  &\quad\qquad\textstyle+\textsf{SNR}_{\mathrm{dl}}^{-1}D_{\mathrm{dl},t}\log_2(D_{\mathrm{dl},t})\left(\frac{2^{\frac{D_{\mathrm{dl},t}}{B(\phi-\tau_{f_k})}}}{B(\phi-\tau_{f_k})^2}\right)\label{eq:T_ul_bound_tmp2}
\end{align}
The $\phi$ that maximizes~\eqref{eq:T_ul_bound_tmp} is given by a root in this equation, which does not have a closed-form solution. Instead we aim to pick $\phi$ such that we obtain a closed-form solution that still provides a good bound. 
To this end, note that \cref{eq:T_ul_bound_tmp2} is dominated by the exponents, and thus a reasonable strategy would be to pick $\phi$ to balance the exponents, i.e., to satisfy $\frac{D_{\mathrm{ul},t}}{B(T-\phi-\tau_{\mathrm{ul},l})}=\frac{D_{\mathrm{dl},t}}{B(\phi-\tau_{f_k})}$.
By isolating $\phi$ we obtain
\begin{align*}
  \phi=\frac{D_{\mathrm{ul},t}\tau_{f_k}+D_{\mathrm{dl},t}(T-\tau_{\mathrm{ul},l})}{D_{\mathrm{ul},t}+D_{\mathrm{dl},t}}.
\end{align*}
Substituting this into~\eqref{eq:T_ul_bound_tmp} yields
\begin{align*}
  &\Pr(T_{\mathrm{tot},t}\le T\,|\,D_{\mathrm{ul},t},D_{\mathrm{dl},t},g_{l,k})\nonumber\\
  &\quad\ge \exp\left(\frac{1-2^{\frac{D_{\mathrm{ul},t}+D_{\mathrm{dl},t}}{B(T-\tau_{\mathrm{ul},l}-\tau_{f_k})}}}{\textsf{SNR}_{\mathrm{ul}}}+\frac{1-2^{\frac{D_{\mathrm{ul},t}+D_{\mathrm{dl},t}}{B(T-\tau_{\mathrm{ul},l}-\tau_{f_k})}}}{\textsf{SNR}_{\mathrm{dl}}}\right)\\
  &\quad= \exp\left(\left(\textsf{SNR}_{\mathrm{ul}}^{-1}+\textsf{SNR}_{\mathrm{dl}}^{-1}\right)\left(1-2^{\frac{D_{\mathrm{ul},t}+D_{\mathrm{dl},t}}{B(T-\tau_{\mathrm{ul},l}-\tau_{f_k})}}\right)\right),
\end{align*}
which is the desired expression.

\section{Proof of \texorpdfstring{\cref{lemma:joint_size_lb}}{Lemma~\ref{lemma:joint_size_lb}}}\label{app:proof:lemma:joint_size_lb}
From Boole's inequality,
\begin{align}
  &\Pr(D_{\mathrm{ul},t}\le \bar{D}_{\mathrm{ul},l}(n),D_{\mathrm{dl},t}\le \bar{D}_{\mathrm{dl},l,k}(m) \,|\, g_{l,k})\nonumber\\
  &\quad\ge \Pr(D_{\mathrm{ul},t}\le \bar{D}_{\mathrm{ul},l}(n) \,|\, g_{l,k})\nonumber\\
  &\quad\quad+ Pr(D_{\mathrm{dl},t}\le \bar{D}_{\mathrm{dl},l,k}(m) \,|\, g_{l,k}) - 1.\label{eq:lemma_marginal_bound_tmp2}
\end{align}

Conditioned on the threshold $\lambda_{l,k}$ and the model choice $g_{l,k}$, the marginal data size samples $\{D_{\mathrm{ul},l}(X_n^{(\mathcal{U})})\}_{n=1}^{N_{\mathcal{U}}}$ and $\{D_{\mathrm{dl},l,k}(X_n^{(\mathcal{U})})\}_{n=1}^{N_{\mathcal{U}}}$ are each a collection of independent samples drawn from the marginal distributions $p(D_{\mathrm{ul},l}(X_n) \,|\, g_{l,k})$ and $p(D_{\mathrm{dl},l,k}(X_n) \,|\, g_{l,k})$, respectively. The data sizes $D_{\mathrm{ul},l}(X)$ and $D_{\mathrm{dl},l,k}(X)$ of a new sample $X\sim P_{X}$ are equally likely to fall in anywhere between the calibration samples, i.e.,
\begin{align*}
  \Pr(D_{\mathrm{ul},l}(X)\le \bar{D}_{\mathrm{ul},l}(n) \,|\, g_{l,k}) &= \frac{n}{N_{\mathcal{U}}+1},\\
  \Pr(D_{\mathrm{dl},l,k}(X)\le \bar{D}_{\mathrm{dl},l,k}(m) \,|\, g_{l,k}) &= \frac{m}{N_{\mathcal{U}}+1},
\end{align*}
for any integers $n,m\in\{1,\ldots,N_{\mathcal{U}}\}$ (see e.g.,~\cite[Appendix D]{conformal}).
Note that this probability is taken over $X\sim P_{X}$, since the data sizes given $X$ and the model choice $g_{l,k}$ are deterministic.

Combining this result with \cref{eq:lemma_marginal_bound_tmp2} yields
\begin{align*}
  &\Pr(D_{\mathrm{ul},t}\le \bar{D}_{\mathrm{ul},l}(n),D_{\mathrm{dl},t}\le \bar{D}_{\mathrm{dl},l,k}(m) \,|\, g_{l,k})\nonumber\\
  &\quad\ge \frac{n}{N_{\mathcal{U}}+1}+\frac{m}{N_{\mathcal{U}}+1}-1\\
  &\quad= \frac{n+m}{N_{\mathcal{U}}+1}-1.
\end{align*}
for any $n,m\in\{1,\ldots,N_{\mathcal{U}}\}$. This completes the proof.

\section{Proof of \texorpdfstring{\cref{prop:delay_violation_dynamic}}{Proposition~\ref{prop:delay_violation_dynamic}}}\label{app:proof:prop:delay_violation_dynamic}
The proof is similar to that of \cref{prop:delay_violation}. We first have the following bound similar to \cref{lemma:conditional_lb}.
\begin{lemma}\label{lemma:conditional_lb_dynamic}
  For a composite model $g_{l,k}$ satisfying $\tau_{\mathrm{ul},l}+\tau_{f_k}\le T$, the conditional probability of satisfying the deadline given $D_{\mathrm{ul},t}$, $D_{\mathrm{dl},t}$, and $R_{\mathrm{ul},t}$ is lower bounded as
  \begin{align*}
    &\Pr(T_{\mathrm{tot},t}\le T \,|\, D_{\mathrm{ul},t},D_{\mathrm{dl},t},R_{\mathrm{ul},t},g_{l,k})\nonumber\\
    &\quad\ge \exp\left(\textsf{SNR}_{\mathrm{dl}}^{-1}\left(1-2^\frac{D_{\mathrm{dl},t}}{B(T-\tau_{\mathrm{ul},l}-\tau_{f_k}-D_{\mathrm{ul},t}/R_{\mathrm{ul},t})}\right)\right).
  \end{align*}
\end{lemma}
\begin{IEEEproof}
  See \cref{app:proof:lemma:conditional_lb_dynamic}.
\end{IEEEproof}
The remainder of the proof proceeds exactly as the proof of \cref{prop:delay_violation}, but using the bound in \cref{lemma:conditional_lb_dynamic} instead of \cref{lemma:conditional_lb}. Specifically, using the fact that $D_{\mathrm{ul},t}$ and $D_{\mathrm{ul},t}$ are independent of $R_{\mathrm{ul},t}$, for any $D_{\mathrm{ul},t}'$ and $D_{\mathrm{dl},t}'$ we have
\begin{align*}
  &\Pr(T_{\mathrm{tot},t} > T \,|\, R_{\mathrm{ul},t},g_{l,k})\nonumber\\
  &\quad\le 1\!-\!\Pr(T_{\mathrm{tot},t}\!\le\! T |D_{\mathrm{ul},t}\!=\!D_{\mathrm{ul},t}',D_{\mathrm{dl},t}\!=\!D_{\mathrm{dl},t}',R_{\mathrm{ul},t},g_{l,k})\nonumber\\
  &\quad\qquad\quad\times \Pr(D_{\mathrm{ul},t}\le D_{\mathrm{ul},t}',D_{\mathrm{dl},t}\le D_{\mathrm{dl},t}' \,|\,g_{l,k})\\
  &\quad\le  1- \hat{L}(D_{\mathrm{ul},t}', D_{\mathrm{dl},t}',R_{\mathrm{ul},t},l,k)\nonumber\\
  &\quad\qquad\quad\times\Pr(D_{\mathrm{ul},t}\le D_{\mathrm{ul},t}',D_{\mathrm{dl},t}\le D_{\mathrm{dl},t}' \,|\, g_{l,k}),
\end{align*}
  where
\begin{align*}
  &\hat{L}(D_{\mathrm{ul},t},D_{\mathrm{dl},t},R_{\mathrm{ul},t},l,k)\nonumber\\
  &\quad=\exp\left(\textsf{SNR}_{\mathrm{dl}}^{-1}\left(1-2^\frac{D_{\mathrm{dl},t}}{B(T-\tau_{\mathrm{ul},l}-\tau_{f_k}-D_{\mathrm{ul},t}/R_{\mathrm{ul},t})}\right)\right).
\end{align*}
Using \cref{lemma:joint_size_lb}, we obtain
  \begin{align*}
  &\Pr(T_{\mathrm{tot},t} > T \,|\, R_{\mathrm{ul},t},g_{l,k})\nonumber\\
  &\quad\le 1- \hat{L}\left(\bar{D}_{\mathrm{ul},l}(n), \bar{D}_{\mathrm{dl},l,k}(m),R_{\mathrm{ul},t},l,k\right)\nonumber\\
  &\quad\qquad\quad\times\left(\frac{n+m}{N_{\mathcal{U}}+1}-1\right)
\end{align*}
for any $n,m\in\{1,\ldots,N_{\mathcal{U}}\}$,
where 
$\bar{D}_{\mathrm{ul},l}(n)$ and $\bar{D}_{\mathrm{dl},l,k}(m)$ are defined as in \cref{eq:def_bar_d_ul_dynamic,eq:def_bar_d_dl_dynamic}, respectively.
The proof is completed by defining
\begin{align*}
  \textstyle\hat{\beta}_{\mathrm{cal}}(l,k,n,m)&=\ln \hat{L}\left(\bar{D}_{\mathrm{ul},l}(n), \bar{D}_{\mathrm{dl},l,k}(m),R_{\mathrm{ul},t},l,k\right)\\
  &=\textstyle\textsf{SNR}_{\mathrm{dl}}^{-1}\!\left(\!1\!-\!2^\frac{\bar{D}_{\mathrm{dl},l,k}(m)}{B(T-\tau_{\mathrm{ul},l}-\tau_{f_k}-\bar{D}_{\mathrm{ul},l}(n)/R_{\mathrm{ul},t})}\right),
\end{align*}
and minimizing over $n$ and $m$.

\section{Proof of \texorpdfstring{\cref{lemma:conditional_lb_dynamic}}{Lemma~\ref{lemma:conditional_lb_dynamic}}}\label{app:proof:lemma:conditional_lb_dynamic}
The proof is similar to that of \cref{lemma:conditional_lb}. For any $0\le \phi\le T$ we have
\begin{align*}
  &\Pr(T_{\mathrm{tot},t}\le T\,|\,D_{\mathrm{ul},t},D_{\mathrm{dl},t},R_{\mathrm{ul},t},g_{l,k})\nonumber\\
  &\quad\ge \Pr(T_{\mathrm{ul},t}\le T-\phi \,|\, D_{\mathrm{ul},t},R_{\mathrm{ul},t},g_{l,k})\nonumber\\
  &\qquad\quad\times\Pr(T_{\mathrm{dl},t}\le \phi\,|\,D_{\mathrm{dl},t},g_{l,k}),
\end{align*}
where we used that $T_{\mathrm{dl},t}$ is independent of $R_{\mathrm{ul},t}$.
Expanding the terms first using \cref{eq:T_ul,eq:T_dl} and then using \cref{eq:capacity_ul,eq:capacity_dl} yields
\begin{align*}
  &\Pr(T_{\mathrm{ul},t}\le T-\phi \,|\, D_{\mathrm{ul},t},R_{\mathrm{ul},t},g_{l,k})\Pr(T_{\mathrm{dl},t}\le \phi\,|\,D_{\mathrm{dl},t},g_{l,k})\\
  &\quad\textstyle= \mathbbm{1}\left[R_{\mathrm{ul},t}\ge \frac{D_{\mathrm{ul},t}}{T-\phi-\tau_{\mathrm{ul},l}}\right]\Pr\left(R_{\mathrm{dl},t}\ge \frac{D_{\mathrm{dl},t}}{\phi-\tau_{f_{k}}} \,\middle|\, D_{\mathrm{dl},t}\right)\\
  &\quad\textstyle=\mathbbm{1}\left[R_{\mathrm{ul},t}\!\ge\! \frac{D_{\mathrm{ul},t}}{T-\phi-\tau_{\mathrm{ul},l}}\right]\Pr\left(\!|h_{\mathrm{dl},t}|^2\!\ge\! \frac{2^{\frac{D_{\mathrm{dl},t}}{B\left(\phi-\tau_{f_{k}}\right)}}-1}{\textsf{SNR}_{\mathrm{dl}}} \middle| D_{\mathrm{dl},t}\!\right)\\
  &\quad\textstyle= \mathbbm{1}\left[R_{\mathrm{ul},t}\ge \frac{D_{\mathrm{ul},t}}{T-\phi-\tau_{\mathrm{ul},l}}\right]
  \exp\left(\frac{1-2^{\frac{D_{\mathrm{dl},t}}{B(\phi-\tau_{f_k})}}}{\textsf{SNR}_{\mathrm{dl}}}\right).
  \end{align*}
  As in the proof of \cref{lemma:conditional_lb}, the best bound is obtained by maximizing the expression over $\phi\in[0,T]$. Since the exponential factor is monotonically increasing in $\phi$, this happens at the largest value of $\phi$ that satisfies the condition in the indicator function, i.e., $\phi= T-\tau_{\mathrm{ul},l}-D_{\mathrm{ul},t}/R_{\mathrm{ul},t}$. By substituting this into the bound, we obtain
  \begin{align*}
  &\Pr(T_{\mathrm{tot},t}\le T\,|\,D_{\mathrm{ul},t},D_{\mathrm{dl},t},R_{\mathrm{ul},t},g_{l,k})\nonumber\\
  &\quad\textstyle\ge
  \exp\left(\frac{1-2^{\frac{D_{\mathrm{dl},t}}{B(T-\tau_{\mathrm{ul},l}-D_{\mathrm{ul},t}/R_{\mathrm{ul},t}-\tau_{f_k})}}}{\textsf{SNR}_{\mathrm{dl}}}\right).
\end{align*}
Rearranging yields the desired result.

\bibliographystyle{IEEEtran}
\bibliography{references}

\begin{thebibliography}{10}
\providecommand{\url}[1]{#1}
\csname url@samestyle\endcsname
\providecommand{\newblock}{\relax}
\providecommand{\bibinfo}[2]{#2}
\providecommand{\BIBentrySTDinterwordspacing}{\spaceskip=0pt\relax}
\providecommand{\BIBentryALTinterwordstretchfactor}{4}
\providecommand{\BIBentryALTinterwordspacing}{\spaceskip=\fontdimen2\font plus
\BIBentryALTinterwordstretchfactor\fontdimen3\font minus
  \fontdimen4\font\relax}
\providecommand{\BIBforeignlanguage}[2]{{%
\expandafter\ifx\csname l@#1\endcsname\relax
\typeout{** WARNING: IEEEtran.bst: No hyphenation pattern has been}%
\typeout{** loaded for the language `#1'. Using the pattern for}%
\typeout{** the default language instead.}%
\else
\language=\csname l@#1\endcsname
\fi
#2}}
\providecommand{\BIBdecl}{\relax}
\BIBdecl

\bibitem{letaief2021edgeai}
K.~B. Letaief, Y.~Shi, J.~Lu, and J.~Lu, ``Edge artificial intelligence for
  {6G}: Vision, enabling technologies, and applications,'' \emph{IEEE J. Sel.
  Areas Commun.}, vol.~40, no.~1, pp. 5--36, 2022.

\bibitem{massiveandcritical24}
A.~E. Kal{\o}r \emph{et~al.}, ``Wireless {6G} connectivity for massive number
  of devices and critical services,'' \emph{Proc. IEEE}, 2024, early access.

\bibitem{zhang20vehicularedge}
J.~Zhang and K.~B. Letaief, ``Mobile edge intelligence and computing for the
  {Internet} of vehicles,'' \emph{Proc. IEEE}, vol. 108, no.~2, pp. 246--261,
  2020.

\bibitem{knowledgebasedurllcinf}
Q.~Zeng, Z.~Wang, Y.~Zhou, H.~Wu, L.~Yang, and K.~Huang, ``Knowledge-based
  ultra-low-latency semantic communications for robotic edge intelligence,''
  \emph{IEEE Trans. Commun.}, 2024, early access.

\bibitem{3gpp_ts22104}
3GPP, ``{Service requirements for cyber-physical control applications in
  vertical domains},'' {3rd Generation Partnership Project (3GPP)}, TS 22.104,
  June 2024, version 19.2.0.

\bibitem{pmlr-v97-tan19a}
M.~Tan and Q.~Le, ``{E}fficient{N}et: Rethinking model scaling for
  convolutional neural networks,'' in \emph{Proc. Int. Conf. Mach. Learn.
  (ICML)}, 2019, pp. 6105--6114.

\bibitem{conformal}
A.~N. Angelopoulos and S.~Bates, ``Conformal prediction: A gentle
  introduction,'' \emph{Found. Trends Mach. Learn}, vol.~16, no.~4, pp.
  494--591, 2023.

\bibitem{conformalrisk}
A.~N. Angelopoulos, S.~Bates, A.~Fisch, L.~Lei, and T.~Schuster, ``Conformal
  risk control,'' in \emph{Proc. Int. Conf. Learn. Represent. (ICLR)}, 2024.

\bibitem{angelopoulosppi}
A.~N. Angelopoulos, S.~Bates, C.~Fannjiang, M.~I. Jordan, and T.~Zrnic,
  ``Prediction-powered inference,'' \emph{Science}, vol. 382, no. 6671, pp.
  669--674, 2023.

\bibitem{shao2020edgeai}
J.~Shao and J.~Zhang, ``Communication-computation trade-off in
  resource-constrained edge inference,'' \emph{IEEE Commun. Mag.}, vol.~58,
  no.~12, pp. 20--26, 2020.

\bibitem{huang2020edge}
X.~Huang and S.~Zhou, ``Dynamic compression ratio selection for edge inference
  systems with hard deadlines,'' \emph{IEEE Internet Things J.}, vol.~7, no.~9,
  pp. 8800--8810, 2020.

\bibitem{Chen-TWC-2019}
E.~Li, L.~Zeng, Z.~Zhou, and X.~Chen, ``Edge {AI}: On-demand accelerating deep
  neural network inference via edge computing,'' \emph{IEEE Trans. Wireless
  Commun.}, vol.~19, no.~1, pp. 447--457, 2019.

\bibitem{li2028jalad}
H.~Li, C.~Hu, J.~Jiang, Z.~Wang, Y.~Wen, and W.~Zhu, ``{JALAD}: Joint
  accuracy-and latency-aware deep structure decoupling for edge-cloud
  execution,'' in \emph{Proc. IEEE Int. Conf. Parallel Dist. Syst. (ICPADS)},
  2018, pp. 671--678.

\bibitem{jankowski2020edgeinf}
M.~Jankowski, D.~Gündüz, and K.~Mikolajczyk, ``Joint device-edge inference
  over wireless links with pruning,'' in \emph{Proc. IEEE Int. Workshp. Signal
  Process. Adv. Wireless Commun. (SPAWC)}, 2020, pp. 1--5.

\bibitem{jankowski2021imgretrieval}
------, ``Wireless image retrieval at the edge,'' \emph{IEEE J. Sel. Areas
  Commun.}, vol.~39, no.~1, pp. 89--100, 2021.

\bibitem{urllc-inference}
Z.~Wang, A.~E. Kal{\o}r, Y.~Zhou, P.~Popovski, and K.~Huang,
  ``Ultra-low-latency edge inference for distributed sensing,''
  \emph{arXiv:2407.13360}, 2024.

\bibitem{urllc-sensing}
Q.~Zeng, J.~Huang, Z.~Wang, K.~Huang, and K.~K. Leung, ``Ultra-low-latency edge
  intelligent sensing: A source-channel tradeoff and its application to coding
  rate adaptation,'' \emph{arXiv:2503.04645}, 2025.

\bibitem{liu23earlyexiting}
Z.~Liu, Q.~Lan, and K.~Huang, ``Resource allocation for multiuser edge
  inference with batching and early exiting,'' \emph{IEEE J. Sel. Areas
  Commun.}, vol.~41, no.~4, pp. 1186--1200, 2023.

\bibitem{she23edgescheduling}
Y.~She, M.~Li, Y.~Jin, M.~Xu, J.~Wang, and B.~Liu, ``On-demand edge inference
  scheduling with accuracy and deadline guarantee,'' in \emph{Proc. IEEE/ACM
  Int. Symp. Qual. Service (IWQoS)}, 2023, pp. 1--10.

\bibitem{jankowski2024earlyexit}
M.~Jankowski, D.~Gündüz, and K.~Mikolajczyk, ``Adaptive early exiting for
  collaborative inference over noisy wireless channels,'' in \emph{Proc. IEEE
  Int. Conf. Mach. Learn. Commun. Netw. (ICMLCN)}, 2024, pp. 126--131.

\bibitem{Zhiyan-AirPooling}
Z.~Liu, Q.~Lan, A.~E. Kalør, P.~Popovski, and K.~Huang, ``Over-the-air
  multi-view pooling for distributed sensing,'' \emph{IEEE Trans. Wireless
  Commun.}, vol.~23, no.~7, pp. 7652--7667, 2024.

\bibitem{lan23progressive}
Q.~Lan, Q.~Zeng, P.~Popovski, D.~Gündüz, and K.~Huang, ``Progressive feature
  transmission for split classification at the wireless edge,'' \emph{IEEE
  Trans. Wireless Commun.}, vol.~22, no.~6, pp. 3837--3852, 2023.

\bibitem{zhang25beyondcloud}
X.~Zhang \emph{et~al.}, ``Beyond the cloud: Edge inference for generative large
  language models in wireless networks,'' \emph{IEEE Trans. Wireless Commun.},
  vol.~24, no.~1, pp. 643--658, 2025.

\bibitem{cohen23conformalurllc}
K.~M. Cohen, S.~Park, O.~Simeone, P.~Popovski, and S.~Shamai, ``Guaranteed
  dynamic scheduling of ultra-reliable low-latency traffic via conformal
  prediction,'' \emph{IEEE Signal Process. Lett.}, vol.~30, pp. 473--477, 2023.

\bibitem{cohen23conformal}
K.~M. Cohen, S.~Park, O.~Simeone, and S.~Shamai~Shitz, ``Calibrating {AI}
  models for wireless communications via conformal prediction,'' \emph{IEEE
  Trans. Mach. Learn. Commun. Netw.}, vol.~1, pp. 296--312, 2023.

\bibitem{zhu24federated}
M.~Zhu, M.~Zecchin, S.~Park, C.~Guo, C.~Feng, and O.~Simeone, ``Federated
  inference with reliable uncertainty quantification over wireless channels via
  conformal prediction,'' \emph{IEEE Trans. Signal Process.}, vol.~72, pp.
  1235--1250, 2024.

\bibitem{earlyexiting}
S.~Teerapittayanon, B.~McDanel, and H.~Kung, ``Branchynet: Fast inference via
  early exiting from deep neural networks,'' in \emph{proc. Int. Conf. Pattern
  Recognit. (ICPR)}, 2016, pp. 2464--2469.

\bibitem{wang2024yolov10}
A.~Wang \emph{et~al.}, ``{YOLOv10}: Real-time end-to-end object detection,''
  \emph{arXiv:2405.14458}, 2024.

\bibitem{Angelopoulos2021LTT}
A.~N. Angelopoulos, S.~Bates, E.~J. Cand{\`e}s, M.~I. Jordan, and L.~Lei,
  ``Learn then test: Calibrating predictive algorithms to achieve risk
  control,'' \emph{arXiv:2110.01052}, 2021.

\bibitem{zecchin2025allt}
M.~Zecchin, S.~Park, and O.~Simeone, ``Adaptive learn-then-test: Statistically
  valid and efficient hyperparameter selection,'' \emph{arXiv:2409.15844},
  2025.

\bibitem{imagenet}
O.~Russakovsky \emph{et~al.}, ``{ImageNet} large scale visual recognition
  challenge,'' \emph{Int. J. Comput. Vision (IJCV)}, vol. 115, no.~3, pp.
  211--252, 2015.

\bibitem{webp}
\BIBentryALTinterwordspacing
J.~Zern, P.~Massimino, and J.~Alakuijala, ``{WebP} image format,'' Nov. 2024.
  [Online]. Available: \url{https://www.rfc-editor.org/info/rfc9649}
\BIBentrySTDinterwordspacing

\bibitem{pmlr-v139-tan21a}
M.~Tan and Q.~Le, ``{EfficientNetV2}: Smaller models and faster training,'' in
  \emph{Proc. Int. Conf. Mach. Learn. (ICML)}, 2021, pp. 10\,096--10\,106.

\bibitem{pytorch}
J.~Ansel \emph{et~al.}, ``{PyTorch} 2: Faster machine learning through dynamic
  {Python} bytecode transformation and graph compilation,'' in \emph{Proc. ACM
  Int. Conf. Architectural Support Program. Lang. Operating Syst., Vol. 2},
  2024, p. 929–947.

\end{thebibliography}

\end{document}